\documentclass[12pt]{article}
\usepackage{latexsym}
\usepackage{bbm}
\usepackage{theorem}
\usepackage{amsopn}
\usepackage{amsfonts}
\usepackage{amssymb}
\usepackage{amsmath}
\usepackage[english]{babel}
\renewcommand{\baselinestretch}{1.2}
\normalsize
\newtheorem{Theorem}{Theorem}[section]
\newtheorem{Proposition}[Theorem]{Proposition}
\newtheorem{Lemma}[Theorem]{Lemma}

\newtheorem{Definition}[Theorem]{Definition}

\theorembodyfont{\rmfamily}

\newenvironment{beweis}{\noindent{\em Proof:}}{\hfill $\rule{2mm}{2mm}$
\vspace{3mm}\par}

\DeclareMathAlphabet{\Ma}{U}{msa}{m}{n}
\DeclareMathAlphabet{\Mb}{U}{msb}{m}{n}
\DeclareMathAlphabet{\Meuf}{U}{euf}{m}{n}

\def\got#1{\Meuf{#1}}

\DeclareSymbolFont{ASMa}{U}{msa}{m}{n}
\DeclareSymbolFont{ASMb}{U}{msb}{m}{n}
\DeclareMathSymbol{\hrist}{\mathord}{ASMa}{"16}
\DeclareMathSymbol{\varkappa}{\mathalpha}{ASMb}{"7B}
\DeclareMathSymbol{\CrPr}{\mathord}{ASMb}{"6F}

\newfont{\EinsFont}{cmr7 scaled 1070}
\def\EINS{{\mathchoice{
 \mbox{\unitlength1cm\begin{picture}(.25,.2)\put(0,0){$1$}%
 \put(0.105,0){{\mbox{\fontfamily{cmr}\upshape\small l}}}\end{picture}}}{%
 \mbox{\unitlength1cm\begin{picture}(.25,.2)\put(0,0){$1$}%
 \put(0.105,0){{\mbox{\fontfamily{cmr}\upshape\small l}}}\end{picture}}}{%
 \mbox{\unitlength1cm\begin{picture}(.18,.15)\put(0,0){$\scriptstyle 1$}%
 \put(0.07,0){{\mbox{\fontfamily{cmr}\upshape\EinsFont l}}}\end{picture}}}{%
 \mbox{\unitlength1cm\begin{picture}(.18,.15)\put(0,0){$\scriptstyle 1$}%
 \put(0.07,0){{\mbox{\fontfamily{cmr}\upshape\EinsFont l}}}\end{picture}}}}}
\def\un{\EINS}
\def\restriction{{\mathchoice{
 \mbox{\unitlength1cm\begin{picture}(.2,.4)%
  \bezier{5}(.07,.3)(.1,.27)(.13,.24)%
  \put(.07,.35){\line(0,-1){.5}}\end{picture}}}{
 \mbox{\unitlength1cm\begin{picture}(.2,.4)%
  \bezier{5}(.07,.3)(.1,.27)(.13,.24)%
  \put(.07,.35){\line(0,-1){.5}}\end{picture}}}{
  \hrist}{\hrist}}}

  \def\al #1.{{\mathcal{#1}}}
  \def\ot #1.{{\got{#1}}}

  \def\ccr #1,#2.{\overline{\Delta(#1,\,#2)}}

  \def\b #1.{{\bf #1}}
  \def\cross#1.{\mathrel{\mathop{\times}\limits_{#1}}}
  \def\tensor#1.{\mathrel{\mathord{\otimes}_{#1}}}
  \def\B{\Theta}
  \def\C{\Mb{C}}
  
  \def\N{\Mb{N}}
  
  \def\R{\Mb{R}}
  
  \def\T{\Mb{T}}
  \def\Z{\Mb{Z}}
  \def\Cn{{\bf V}}

  \def\wt{\widetilde}
\def\ilim{\mathop{{\rm lim}}\limits_{\longrightarrow}}
\def\cross #1.{\mathrel{\raise 3pt\hbox{$\mathop\times\limits_{#1}$}}}
\def\crossr #1.{\mathrel{\raise 1pt\hbox{$\mathop\rtimes\limits_{#1}$}}}
\def\maprightu #1;{\mathrel{\smash{\mathop{\longrightarrow}\limits^{#1}}}}
\def\maprightb #1;#2.{\mathrel{\smash{\mathop{\longrightarrow}\limits_{#2}^{#1}}}}
\def\s #1.{_{\smash{\lower2pt\hbox{\mathsurround=0pt $\scriptstyle #1$}}\mathsurround=5pt}}
\def\cl #1.{{\cal #1}} \def\al #1.{{\cal #1}}
\def\ker{{\rm Ker}\,}
\def\aut{{\rm Aut}\,}
\def\gau{{\rm Gau}\,}
\def\gaud{{\rm Gau}_d\,}
\def\gaue{{\rm Gau}^e\,}
\def\gaued{{\rm Gau}_d^e\,}

\def\Lgau{{\got{gau}}\,}

\def\gauc{{\rm Gau}\,}

\def\f #1,#2.{\mathsurround=0pt \hbox{${#1\over #2}$}\mathsurround=5pt}

\def\ol#1.{\overline{#1}}

\def\CAR #1.{{{\rm CAR}(#1)}}

\def\fl #1,#2.{{\mathord{\rm Fl}^{#1}_{#2}}}

\def\XP#1!{\renewcommand{\baselinestretch}{.7}\marginpar{{\footnotesize
$\leftarrow$#1}\hfil}\renewcommand{\baselinestretch}{1.5}}
\def\f #1,#2.{\mathsurround=0pt \hbox{${#1\over #2}$}\mathsurround=5pt}
   
\def\set #1,#2.{\left\{\,#1\;\bigm|\;#2\,\right\}}

\newcommand{\Aut}{\mathop{{\rm Aut}}\nolimits}

\newcommand{\into}{\hookrightarrow}

\newcommand{\supp}{\mathop{{\rm supp}}\nolimits}

\renewcommand{\hat}{\widehat}
\renewcommand{\tilde}{\widetilde}

\newcommand{\cH}{\mathcal{H}}

\newcommand{\la}{\langle}
\newcommand{\ra}{\rangle}
\newcommand{\1}{\mathbf{1}}

\def\slim{\mathop{\hbox{\rm s-lim}}}
\def\bn{{\bf n}}

\begin{document}


\title{\bf Dynamics for QCD on an infinite lattice}

\author{
  {\sc Hendrik Grundling}                                            \\[1mm]
 {\footnotesize Department of Mathematics,}                         \\
 {\footnotesize University of New South Wales,}                      \\
 {\footnotesize Sydney, NSW 2052, Australia.}                             \\
 {\footnotesize hendrik@maths.unsw.edu.au}                  \\
 {\footnotesize FAX: +61-2-93857123} \\
\and
 {\sc  Gerd Rudolph }                                            \\[1mm]
 {\footnotesize Institut f\"ur Theoretische Physik,}                         \\
 {\footnotesize Universit\"at Leipzig,}                      \\
 {\footnotesize Postfach 100 920, D-4109 Leipzig.}                             \\
 {\footnotesize rudolph@rz.uni-leipzig.de}                  \\
 {\footnotesize FAX: +49-341-9732548}}

\maketitle


\begin{abstract}
We prove the existence of the dynamics automorphism group for Hamiltonian
QCD on an infinite lattice in $\R^3$, and this is done in a C*-algebraic context.
The existence of ground states is also obtained.
Starting with the finite lattice model for Hamiltonian QCD developed by
Kijowski, Rudolph  (cf.~\cite{KR, KR1}), we state its field algebra
and a natural representation. We then generalize this representation to the infinite lattice,
and construct a Hilbert space which has represented on it all the local algebras
(i.e. kinematics algebras associated with finite connected sublattices) equipped with the correct
graded commutation relations. On a suitably large C*-algebra acting on this Hilbert space, and
containing all the local algebras, we prove that there is a one parameter automorphism group, which is the
pointwise norm limit of the local time evolutions along a sequence of finite sublattices, increasing to the full lattice.
This is our global time evolution. We then take as our field algebra the C*-algebra generated by
all the orbits of the local algebras w.r.t. the global time evolution.
Thus the time evolution creates the field algebra.
The time evolution is strongly continuous
 on this choice of field algebra, though not on the original larger C*-algebra.
We define the gauge transformations, explain how
to enforce
the Gauss law constraint, show that the dynamics automorphism group
descends to the algebra of physical observables
and prove that gauge invariant ground states exist.
\end{abstract}

\newpage


\setcounter{equation}{0}
\section{Introduction}



In a previous paper (\cite{GrRu}) we constructed in a C*-algebraic context a suitable field algebra
which can model the kinematics of Hamiltonian
QCD on an infinite lattice in $\R^3$.
It was based on the finite lattice model for Hamiltonian QCD developed by
Kijowski, Rudolph  (cf.~\cite{KR, KR1}).
We did not consider dynamics, and the construction and analysis of the dynamics for
QCD on an infinite lattice is the main problem which we want to address in this paper.
For reasons to be explained, we will not here directly
use the C*-algebra which we constructed before for dynamics construction, but will
follow a different approach.

Whilst the algebra constructed in \cite{GrRu} contained all the information of the gauge structures required,
and its representation space contained the physical representations, it suffered from
the following defects.
\begin{itemize}
\item{}
The true local algebras (i.e. the kinematics algebras for the model on finite sublattices)
 were not subalgebras of the constructed kinematics algebra.
 The kinematics algebra of \cite{GrRu} did contain isomorphic copies of the local algebras, whose
 multiplier algebras contained the local algebras, but these did not satisfy local graded-commutativity.
That is, they did not graded-commute if they corresponded to disjoint parts of the lattice, unlike the true local algebras.
 This was due to a novel form of the
infinite tensor product of nonunital algebras,
where approximate identities replaced the identity in the infinite ``tails" of the tensor products.
\item{}
As a consequence of this infinite tensor product,
current methods of defining dynamics on lattice systems by suitable limits of
the local dynamics did not apply, which made it very hard to construct dynamics for the full system.
\end{itemize}
 Due to these problems, especially the latter one,
 we will here extend our focus to the multiplier algebra of our previous kinematics field algebra,
 and build an appropriate new kinematics field algebra in this setting.
Our strategy will be to define a dynamics on a concrete C*-algebra which is ``maximally large'' in the sense that
 it contains all the true local kinematics C*-algebras, and it is contained in
 the multiplier algebra of our previous kinematics field algebra.
  We will then take our new kinematics field C*-algebra to be the smallest subalgebra
 which contains  all the true local kinematics C*-algebras, and is preserved w.r.t. the dynamics.
 Thus, the dynamics itself, will create the kinematics algebra for the system.
 The methods we use for the proof come from the application and generalization of Lieb--Robinson bounds
 on lattice systems (cf.~\cite{NaSi, NaSi2}).
 Moreover, we will see that the dynamics is strongly continuous on our new kinematics algebra.

 On the algebra we construct here, we are able to define both the dynamics and the gauge transformations
 of our model. We will prove that
the dynamics automorphism group commutes with the action of the group of
local gauge transformations, hence the dynamics automorphism group action
descends to the algebra of physical observables.
We will prove  the existence of gauge invariant ground states, which  therefore produces
ground states on the algebra of physical observables.

The cost of using this new algebra, is that it  contains infinitely many nonphysical representations,
 so one needs to restrict to the class of appropriate ``regular'' representations by hand.
 This should be compared with the use of the Weyl algebra for canonical systems, which contains many nonregular representations.
 It is a well-known practical necessity for the Weyl algebra to restrict representations
  by hand to regular representations when one analyzes physical systems.
  We are able to prove the existence of gauge invariant ground states which are regular.


 Whereas each true local kinematics algebra (corresponding to a finite sublattice)
has a unique ground state w.r.t. the local time evolution,
for the infinite lattice limit,
we do not presently have such a uniqueness property.
The  ground states are the
weak *-limit points of a sequence of ``partial ground states''. This
nonuniqueness needs further investigation.

For the finite lattice, in a
 toy model
  the spectral problem for the lattice Hamiltonian has
been solved exactly, yielding  an explicit formula for the
unique ground state, cf.~\cite{HRS}.

Our paper is organized as follows. In Sect.~\ref{PM} we state the model
for the finite lattice taken from \cite{KR, KR1}, and give a very natural representation for it.
We then generalize this representation for the infinite lattice in Sect.~\ref{GFA},
and construct a Hilbert space $\cl H.$ which has represented on it all the local algebras ${\mathfrak A}_S$
(each associated with a finite sublattice $S$),
with the correct (graded) commutation relations. We  then define a conveniently large C*-algebra
$\al A._{\rm max}$ acting on this Hilbert space, and containing all the local algebras.
In Sect.~\ref{LDA}, we
then define on $\al A._{\rm max}$  the ``local automorphism groups" $\alpha_t^S$, i.e. those produced by the
Hamiltonians of the finite sublattices $S$. Using a Lieb--Robinson bounds argument, we then prove
in Sect.~\ref{DMFA} that for each $A\in \al A._{\rm max}$, that $\alpha_t^S(A)$ converges in norm as $S$ increases
to the full lattice, to an element $\alpha_t(A)$, and  this defines a one-parameter automorphism group
$t\mapsto\alpha_t\in{\rm Aut}(\al A._{\rm max})$. This is the global automorphism group, and we use it to define
in Sect.~\ref{KARR} our
chosen minimal field algebra by
\[
{\mathfrak A}_{\Lambda}:=C^*\Big(  \bigcup_{S\in\al S.}\alpha_{\R}({\mathfrak A}_S   )    \Big)
\subset\cl A._{\rm max}    \subset\cl B.(\cl H.).
\]
We also clarify the relation of ${\mathfrak A}_{\Lambda}$ with
the kinematics algebra previously constructed in \cite{GrRu}.
We prove the existence of regular ground states in Sect.~\ref{PGDA}, and in Sect.~\ref{GTGL}
we define gauge transformations and consider enforcement of the Gauss law constraint.


\setcounter{equation}{0}
\section{The Kinematics Field Algebra}
\label{FieldAlgebra}

%
%

We consider a model for QCD in the Hamiltonian framework on an infinite regular
cubic lattice in $\Z^3.$
For basic notions concerning lattice gauge
theories including fermions, we refer to \cite{Seiler} and
references therein.

We first fix notation.
For the lattice, define a triple $\Lambda:=(\Lambda^0, \Lambda^1, \Lambda^2)$ as follows:
\begin{itemize}
\item{}
 $\Lambda^0:=\{(n,m,r)\in\R^3\,\mid\,n,\,m,\,r\in\Z\}=\Z^3$
i.e. $\Lambda^0$ is the unit cubic lattice
and its elements are called sites.
\item{} Let $\wt\Lambda^1$ be the set of all directed edges (or links) between nearest neighbours, i.e.
\[
\wt\Lambda^1:= \{(x,y)\in\Lambda^0\times\Lambda^0\,\mid\,y=x\pm \b e._i\;\;\hbox{for some $i$}\}
\]
where the $\b e._i\in\R^3$ are the standard unit basis vectors.
Let
$\Lambda^1\subset\wt\Lambda^1 $ denote a choice of orientation of $\wt\Lambda^1,$
i.e. for each
$(x,y)\in\wt\Lambda^1$,  $\Lambda^1$ contains either $(x,y)$ or $(y,x)$ but not both.
%
Thus the pair $(\Lambda^0,\Lambda^1)$
is a directed graph, and we assume that it is connected.
\item{} Let $\wt\Lambda^2$ be the set of all directed faces (or plaquettes) of the unit cubes comprising the lattice i.e.
\[
\wt\Lambda^2:= \{(\ell_1,\ell_2,\ell_3,\ell_4)\in\big(\wt\Lambda^1\big)^4\,\mid\,Q_2\ell_i=Q_1\ell_{i+1}\;
\hbox{for}\;i=1,2,3,\;\hbox{and}\; Q_2\ell_4=Q_1\ell_1\}
\]
where $Q_i:\Lambda^0\times\Lambda^0\to\Lambda^0$ is the projection onto the $i^{\hbox{th}}$  component.
Note that for a plaquette $p=(\ell_1,\ell_2,\ell_3,\ell_4)\in\wt\Lambda^2,$ it has an orientation given by the order of the edges,
and the  reverse ordering is
$\overline{p}=(\overline\ell_4,\overline\ell_3,\overline\ell_2,\overline\ell_1)$ where
$\overline\ell={(y,x)}$ if $\ell={(x,y)}\in \wt\Lambda^1$.
In analogy to the last point, we let
$\Lambda^2$ be a choice of orientation in $\wt\Lambda^2$.
\item{} If we need to identify the elements of $\Lambda^i$ with subsets of $\R^3$, we will make the natural
identifications, e.g. a link $\ell=(x,y)\in\Lambda^1$ is the undirected closed line segment from $x$ to $y$.
\item{} We also need to consider subsets of the lattice, so
given a connected subgraph $S\subset(\Lambda^0, \Lambda^1)$, we let
$\Lambda_S^0$ be all the vertices in $S$, $\Lambda_S^1\subset\Lambda^1$ is the set of links which are edges in $S$, and
$\Lambda_S^2\subset\Lambda^2$ is the set of those plaquettes whose sides are all in $S$.
\end{itemize}

We recall the lattice approximation on $\Lambda$
for a classical matter field  with a classical gauge
connection field acting on it.
Fix a connected, compact Lie group $G$ (the gauge structure group),
and let ${\big(\Cn,\,(\cdot,\cdot)_\Cn\big)}$ be a finite dimensional complex
Hilbert space  (the space of internal degrees of freedom of the matter field)
on which $G$ acts smoothly as unitaries, so we take $G\subset U(\Cn)$.
Then the classical matter fields are elements of $\prod\limits_{x\in\Lambda^0}\Cn$,
on which the local gauge group $\prod\limits_{x\in \Lambda^0}G=G^{\Lambda^0}=\big\{\zeta:\Lambda^0\to G\big\}$
acts by pointwise multiplication.
The classical gauge connections are maps  $\Phi:\Lambda^1\to G$, i.e. elements of $\prod\limits_{\ell\in\Lambda^1}G$.

The full classical configuration space is thus $\big(\prod\limits_{x\in\Lambda^0}\Cn\big)\times
\big(\prod\limits_{\ell\in\Lambda^1}G\big)$, and the local gauge group $\prod\limits_{x\in \Lambda^0}G$ acts on it  by
\begin{equation}
\label{ClassGTr}
\big(\prod_{x\in\Lambda^0}v_x\big)\times\big(\prod_{\ell\in\Lambda^1}g_\ell\big)\mapsto
\big(\prod_{x\in\Lambda^0}\zeta(x)\cdot v_x\big)\times\big(\prod_{\ell\in\Lambda^1}\zeta(x_\ell)\,g_\ell\,\zeta(y_\ell)^{-1}\big)
\end{equation}
where  $ \ell=(x_\ell,y_\ell)$ and $\zeta \in\prod\limits_{x\in \Lambda^0}G$.
Note that the  orientation of links in $\Lambda^1$ was used in the action
because it treats the $x_\ell$ and $y_\ell$ differently.

This is the basic classical kinematical model for which  the quantum counterpart is given below for finite lattices.


\subsection{The finite lattice model.}
\label{PM}

In this subsection we want to state the model for finite lattice approximation
of Hamiltonian QCD in $\R^3$ developed by Kijowski, Rudolph~\cite{KR, KR1}.
A more expanded version of this section is in \cite{GrRu}.
This model is based on the model constructed in the classical paper of
Kogut~\cite{K} (which elaborates the earlier one of Kogut and Susskind~\cite{KS}).

Fix a  finite connected subgraph $S$, and let
 $\Lambda_S:=(\Lambda^0_S, \Lambda^1_S, \Lambda^2_S)$. For ease of notation, we will omit the subscript $S$ in this section.
Given such a finite lattice $\Lambda^0$, the model quantizes the classical model on  $\Lambda^0$ above, by replacing
for each lattice site $x\in\Lambda^0$, the classical matter configuration space $\Cn$ with
the algebra for a fermionic particle on $\Cn$
(the quarks), and for each link $\ell\in\Lambda^1$ we replace the classical connection configuration space
$G$ by an algebra which describes
a bosonic particle on $G$ (the gluons).

Equip the space of classical matter fields $\prod\limits_{x\in\Lambda^0}\Cn=\{f:\Lambda^0\to\Cn\}$ with the natural pointwise inner product
$\la f, h \ra=\sum\limits_{x\in\Lambda^0}\big({f(x)},\, h(x)\big)_{\Cn}$, and take for the quantized
matter fields the CAR-algebra ${\mathfrak F}_{\Lambda} :=  {\rm CAR}\big(\prod\limits_{x\in\Lambda^0}\Cn)$.
That is, for each classical matter field $f\in\prod\limits_{x\in\Lambda^0}\Cn$,
we associate a fermionic field $a(f)\in{\mathfrak F}_{\Lambda}$, and these
satisfy the usual CAR--relations:
\[
\{a(f), a(h)^*\}  = \la f, h \ra \1
\quad \hbox{ and } \quad  \{a(f),a(h)\} = 0
\quad \mbox{ for } \quad f, h\in \prod\limits_{x\in\Lambda^0}\Cn
\]
where  $\{A,B\} := AB + BA$ and ${\mathfrak F}_{\Lambda}$ is generated by the set of all $a(f)$.
As $\Lambda^0$ is finite,
${\mathfrak F}_{\Lambda}$ is a full matrix algebra, hence up to unitary equivalence
it has only one irreducible representation.

In physics notation,  the quark at $x$ is given by
$a(\delta_x{\bf v}_i)=\psi_i(x)$ where ${\{{\bf v}_i\mid i=1,\ldots n\}}$ is an orthonormal basis for $\Cn$
and $\delta_x:\Lambda^0\to\R$ is the characteristic function of $\{x\}$.
Further indices may be included if necessary, e.g. by putting
$\Cn = \b W.\otimes\C^k$ where $\b W.$ has  non--gauge degrees of freedom (such as the spinor part),
 and $\C^k$ has the gauge degrees of freedom.

To quantize the classical gauge connection fields $\prod\limits_{\ell\in\Lambda^1}G,$
we take for a single link $\ell$  a  bosonic particle on $G$. This is given
in a generalized Schr\"odinger representation on $L^2(G)$
 by the set of operators ${\{U_g,\;T_f\mid g\in G,\;f\in L^\infty(G)\}}$ where:
\begin{equation}
\label{GSRp}
(U_g\varphi)(h):=\varphi(g^{-1}h)\quad\hbox{and}\quad\big(T_f\varphi)(h):=f(h)\varphi(h)\quad\hbox{for}\quad
\varphi\in L^2(G),
\end{equation}
$g,\,h\in G$ and $f\in L^\infty(G)$, and it is irreducible in the sense that the commutant
of ${U_G\cup T_{L^\infty(G)}}$ consists of the scalars.
Note that there is a natural ground state unit vector $\psi_0\in L^2(G)$ given by the constant
function $\psi_0(h)=1$ for all $h\in G$ (assuming that the Haar measure of $G$ is normalized).
 Then $U_g\psi_0=\psi_0$, and $\psi_0$
is cyclic w.r.t. the *-algebra generated by ${U_G\cup T_{L^\infty(G)}}$ (by irreducibility).

 The generalized canonical commutation relations are obtained from the
intertwining relation $U_gT_fU^*_g=T\s\lambda_g(f).$ where
\begin{equation}
\label{leftact}
\lambda:G\to\aut C(G)\, , \quad
 \lambda_g(f)(h) := f(g^{-1} h) \quad\hbox{for}\quad g,\,h\in G
\end{equation}
is the usual left translation. In particular, given $X\in\mathfrak{g}$, define
its associated momentum operator
\begin{eqnarray*}
P_X:C^\infty(G)\to C^\infty(G)\quad&\hbox{by}&\quad
P_X\varphi:=i{d\over dt}U(e^{tX})\varphi\Big|_{t=0}\,.\\[1mm]
\hbox{Then}\qquad
\big[P_X,\,T_f\big]\varphi
&=& iT\s X^R(f). \varphi\qquad\hbox{for}\quad
f,\,\varphi\in C^\infty(G),
\end{eqnarray*}
where $X^R\in\ot X.(G)$ is the associated right-invariant vector field.
As $P_X=dU(X)$, it defines  a representation of the Lie algebra $\mathfrak{g}$,
and clearly $P_X\psi_0=0$.

To identify the quantum connection $\Phi(\ell)$ at link $\ell$ in this context,
use the irreducible action of the structure group $G$ on $\C^k$ to define the function $\Phi_{ij}(\ell)\in C(G)$ by
\begin{equation}
\label{Def-QuConn}
\Phi_{ij}(\ell)(g):=(e_i,ge_j),\quad g\in G,
\end{equation}
 where $\{ e_i\mid
i =1,\ldots,k\}$ is an orthonormal basis of $\C^k$. Then  the matrix components of the quantum connection are
taken to be the operators $T_{\Phi_{ij}(\ell)}$,
which we will see transform correctly w.r.t. gauge transformations.
As the $\Phi_{ij}(\ell)$ are matrix elements of elements of $G$, there are obvious
relations between them which reflect the structure of $G$.
The C*-algebra generated by the operators $\{ T_{\Phi_{ij}(\ell)}\,\mid\, i,\,j=1,\ldots,k\}$
is $T\s C(G).$.

To define gauge momentum operators, we first assign to each link an element of $\mathfrak{g}$,
i.e. we choose a map $\Psi:\Lambda^1\to \mathfrak{g}.$ Given such a $\Psi$,  take
for the associated quantum gauge momentum at $\ell$ the operator
$P\s\Psi(\ell).:C^\infty(G)\to C^\infty(G)$.
 The generalized canonical commutation relations are
\begin{equation}
\label{genCCR}
\big[P\s\Psi(\ell).,\,T\s\Phi_{ij}(\ell).\big]=i\sum_mT\s\Psi(\ell)_{im}\Phi_{mj}(\ell).\quad\hbox{on}\quad C^\infty(G),
\end{equation}
where $\Psi(\ell)_{im}:=(e_i,\Psi(\ell)e_m).$

To obtain the G--electrical fields at $\ell$,  choose a basis
${\{Y_r\mid r=1,\ldots,{\rm dim}( \mathfrak{g})\}\subset \mathfrak{g}}$,
then substitute for $\Psi$ the constant map $\Psi(\ell)=Y_r$ and set
$E_r(\ell):=P_{Y_r}$. In the case that $G=SU(3)$, these are the colour electrical fields,
and one takes  the basis $\{Y_r\}$ to be the traceless selfadjoint Gell--Mann matrices
satisfying ${\rm Tr}(Y_rY_s)=\delta_{rs}$. We then define
\[
E_{ij}(\ell):=\sum_r(Y_r)_{ij}E_r(\ell)=\sum_r(Y_r)_{ij}P\s{Y_r}.
\]
and for these we obtain  from $(\ref{genCCR})$
the commutation formulae in \cite{KR, KR1} for the colour electrical field.
Of particular importance for the dynamics, is the operator $E_{ij}(\ell) E_{ji}(\ell)$ (summation convention). We have
\[
E_{ij}(\ell) E_{ji}(\ell)=\sum_{r,s}(Y_r)_{ij}P\s{Y_r}.(Y_s)_{ji}P\s{Y_s}.
=\sum_{r,s}(Y_rY_s)_{ii}P\s{Y_r}.P\s{Y_s}.=n\sum_rP\s{Y_r}.^2
\]
i.e. it is the Laplacian for the left regular representation $U:G\to\cl U.(L^2(G))$ which therefore commutes with all
$U_g$. Below in Equation~(\ref{PTfell}) we will see that a gauge transform just transforms the Laplacian
to one w.r.t. a transformed basis of $\mathfrak{g}$, which leaves the Laplacian invariant.

The full collection of operators which comprises the set of dynamical
variables of the model is as follows. The representation Hilbert space is
\[
\cl H.=\cl H._F\otimes\mathop{\bigotimes}\limits_{\ell\in\Lambda^1}L^2(G)\qquad\hbox{ where}\qquad
{\pi_F:{\mathfrak F}_{\Lambda}}\to\cl B.(\cl H._F)
\]
 is any
irreducible representation of ${\mathfrak F}_{\Lambda}$.
As $\Lambda^1$ is finite, $\cl H.$ is well--defined. Then
$\pi_F\otimes\un:{\mathfrak F}_{\Lambda}\to\cl B.(\cl H.)$ will be the action of ${\mathfrak F}_{\Lambda}$
on $\cl H.$. The quantum connection is given by the set of operators
\[
\{\hat{T}_{\Phi_{ij}(\ell)}^{(\ell)}\,\mid\,\ell\in \Lambda^1,\;i,j=1,\ldots, k\}\quad
 \hbox{where}\quad \hat{T}_f^{(\ell)}:=\un\otimes\big(\un\otimes\cdots\otimes\un\otimes T^{(\ell)}_f\otimes\un\otimes\cdots\otimes\un\big)
\]
and $T^{(\ell)}_f$ is the multiplication operator on the $\ell^{\rm th}$ factor, hence $\hat{T}_{\Phi_{ij}(\ell)}^{(\ell)}$
acts as the identity on all the other factors of $\cl H.$. Likewise, for the gauge momenta we take
\[
\hat{P}^{(\ell)}_{\Psi(\ell)}:=\un\otimes\big(\un\otimes\cdots\otimes\un\otimes P^{(\ell)}_{\Psi(\ell)}\otimes\un\otimes\cdots\otimes\un\big),\quad
\ell\in \Lambda^1
\]
where $P^{(\ell)}_X$ is the $P_X$ operator on the subspace $C^\infty(G)\subset L^2(G)$ of the $\ell^{\rm th}$ factor.
Note that if we set  $g=\exp(t\Psi(\ell))$ in
 $\hat{U}_g^{(\ell)}:=\un\otimes\big(\un\otimes\cdots\otimes\un\otimes{U}_g^{(\ell)}\otimes\un\otimes\cdots\otimes\un\big)$
where $U^{(\ell)}_g$ is the $U_g$ operator on the $\ell^{\rm th}$ factor,
then this is a unitary one parameter group w.r.t. $t$, with generator the gauge momentum operator
$\hat{P}^{(\ell)}_{\Psi(\ell)}$.
Thus the quantum G--electrical field $\hat{E}_r$ is a map from  $\Lambda^1$ to operators on
the dense domain $\cl H._F\otimes\mathop{\bigotimes}\limits_{\ell\in\Lambda^1}C^\infty(G)$, given by
$\hat{E}_r(\ell):=\hat{P}^{(\ell)}_{Y_r}$.

Next, to define gauge transformations, recall from Equation~(\ref{ClassGTr}) the action of
the local gauge group $\gauc \Lambda=\prod\limits_{x\in \Lambda^0}G=\{\zeta:\Lambda^0\to G\}$ on the
classical configuration space.
For the Fermion algebra we define an action
$\alpha^1:\gauc \Lambda \to\aut{\mathfrak F}_{\Lambda}$ by
\[
\alpha_\zeta^1(a(f)):=a(\zeta\cdot f)\qquad \hbox{where}\qquad (\zeta\cdot f)(x):={\zeta(x)}f(x)\quad \hbox{for all}\quad x\in\Lambda^0,
\]
and $f\in\prod\limits_{x\in\Lambda^0}\Cn$ since $f\mapsto \zeta\cdot f$ defines a unitary on $\prod\limits_{x\in\Lambda^0}\Cn$
where  $\zeta\in\gauc \Lambda$. As ${\mathfrak F}_{\Lambda}$ has up to unitary equivalence
only one irreducible representation, it follows that ${\pi_F:{\mathfrak F}_{\Lambda}}\to\cl B.(\cl H._F)$ is
equivalent to the Fock representation, hence it is covariant w.r.t. $\alpha^1$, i.e. there is a
(continuous) unitary representation $U^F:\gauc \Lambda\to \cl U.(\cl H._F)$ such that
\[
\pi_F(\alpha^1_\zeta(A))=U^F_\zeta\pi_F(A)U^F_{\zeta^{-1}}\quad\hbox{for}\quad A\in {\mathfrak F}_{\Lambda}.
\]

On the other hand,
if the classical configuration space  $G$ corresponds to
a link  $\ell=(x_\ell,y_\ell)$, then the gauge transformation is $\zeta\cdot g=
\zeta(x_\ell)\,g\,\zeta(y_\ell)^{-1}$ for all $g\in G$. Using this, we
define a unitary $W_\zeta:L^2(G)\to L^2(G)$ by
\[
(W_\zeta\varphi)(h):=\varphi(\zeta^{-1}\cdot h)=\varphi(\zeta(x_\ell)^{-1}\,h\,\zeta(y_\ell))
\]
 using the fact that $G$
is unimodular, where the inverse was introduced to ensure that $\zeta\to W_\zeta$ is a homomorphism.
Note that $W_\zeta\psi_0=\psi_0$.
So for the
quantum observables ${U_G\cup T_{L^\infty(G)}}$, the gauge transformation becomes
\begin{equation}
\label{GTfell}
T_f\mapsto W_\zeta T_f W_\zeta^{-1}=T_{W_\zeta f}\quad\hbox{and}\quad
U_g\mapsto W_\zeta U_g W_\zeta^{-1}=U_{\zeta(x_\ell)g\zeta(x_\ell)^{-1}}
\end{equation}
for $f\in L^\infty(G)\subset L^2(G)$ and $g\in G$. Moreover each $W_\zeta$ preserves
the space $C^\infty(G)$, hence Equation~(\ref{GTfell}) also implies that
\begin{equation}
\label{PTfell}
W_\zeta P_X W_\zeta^{-1}=P\s{\zeta(x_\ell)X\zeta(x_\ell)^{-1}}.\quad\hbox{for}\quad
 X\in\mathfrak{g}.
\end{equation}

Thus for the full system we define on $\cl H.=\cl H._F\otimes\mathop{\bigotimes}\limits_{\ell\in\Lambda^1}L^2(G)$
the unitaries
\begin{equation}
\label{LocalW}
\hat{W}_\zeta:=U^F_\zeta\otimes\big(\bigotimes_{\ell\in\Lambda^1}W^{(\ell)}_\zeta\big),\quad
\zeta\in\gauc \Lambda
\end{equation}
where $W^{(\ell)}_\zeta$ is the $W_\zeta$ operator on the $\ell^{\rm th}$ factor. Then the gauge transformation
produced by $\zeta$ on the system of operators is given by ${\rm Ad}(\hat{W}_\zeta)$.

In particular, recalling $W_\zeta T\s\Phi_{ij}(\ell). W_\zeta^{-1}=T\s{W_\zeta \Phi_{ij}(\ell)}.$ we see that
\begin{eqnarray}
\big(W_\zeta \Phi_{ij}(\ell)\big)(g)&=&\Phi_{ij}(\ell)(\zeta(x_\ell)^{-1}\,g\,\zeta(y_\ell))
=\big(e_i,\zeta(x_\ell)^{-1}\,g\,\zeta(y_\ell)e_j\big)\nonumber\\[1mm]
\label{PhiTfs}
&=&\sum_{n,m}[\zeta(x_\ell)^{-1}]_{in}\,\Phi_{nm}(\ell)(g)\,[\zeta(y_\ell)]_{mj}
\end{eqnarray}
where $[\zeta(x_\ell)]_{in}=(e_i,\zeta(x_\ell)e_n)$ are the usual matrix elements, so
it is clear that the indices of the quantum connection $T_{\Phi_{ij}(\ell)}$ transform correctly
for the gauge transformation $\zeta^{-1}$.

Finally, we construct the appropriate field C*-algebra for this model.
For the fermion part, we already have the C*-algebra ${\mathfrak F}_{\Lambda} =  {\rm CAR}\big(\prod\limits_{x\in\Lambda^0}\Cn)$.
Fix a link $\ell$, hence a specific copy of $G$ in the configuration space.
Above in (\ref{leftact}) we had the distinguished action
 $\lambda:G\to\aut C(G)$ by
\[
\lambda_g(f)(h) := f(g^{-1} h) \, \, , \, \,
f \in C(G),\; g,h\in G.
\]
The generalized Schr\"odinger representation $(T,U)$ above in~(\ref{GSRp}) is a covariant representation
for the action $\lambda:G\to\aut C(G)$ so it is natural to take for our field algebra
 the crossed product C*-algebra $C(G)\rtimes_\lambda G $ whose representations are exactly the
 covariant representations of the $C^*$-dynamical system defined by $\lambda$.
The algebra $C(G)\rtimes_\lambda G $ is also called the generalised Weyl algebra, and it
 is well--known that
$C(G)\rtimes_\lambda G\cong\cl K.\big(L^2(G)\big)$ cf.~\cite{Rief} and
Theorem~II.10.4.3 in~\cite{Bla1}.
In fact $\pi_0\big(C(G)\rtimes_\lambda G\big)=\cl K.\big(L^2(G)\big)$
where $\pi_0:C(G)\rtimes_\lambda G\to\cl B.( L^2(G))$ is the generalized Schr\"odinger representation.
Since the algebra of compacts $\cl K.\big(L^2(G)\big)$
has only one irreducible representation up to unitary equivalence, it follows that
the generalized Schr\"odinger representation is the unique irreducible
covariant representation of $\lambda$ (up to equivalence). Moreover, as $\psi_0$ is cyclic for
$\cl K.\big(L^2(G)\big)$,  the
generalized Schr\"odinger representation is unitary equivalent to the GNS--representation
of the vector state $\omega_0$ given by $\omega_0(A):={(\psi_0,\pi_0(A)\psi_0)}$ for $A\in
C(G)\rtimes_\lambda G$.

Note that the operators $U_g$ and $T_f$
in equation~(\ref{GSRp}) are not compact, so they are not in $\cl K.\big(L^2(G)\big)=\pi_0\big(C(G)\rtimes_\lambda G\big)$,
but are in fact in its multiplier algebra. This is not a problem, as
a state or representation on $C(G)\rtimes_\lambda G$ has a unique extension to its multiplier algebra,
so it is fully determined on these elements.
If one chose $C^*(U_G\cup T_{L^\infty(G)})$ as the field algebra instead of
$C(G)\rtimes_\lambda G$, then this would contain many inappropriate representations, e.g.
 covariant representations
for $\lambda:G\to\aut C(G)$ where the implementing unitaries are discontinuous w.r.t. $G$.
Thus, our choice for the field algebra of a link remains as ${C(G)\rtimes_\lambda G}\cong\cl K.\big(L^2(G)\big)$.
Clearly, as the momentum operators $P_X$ are unbounded, they cannot be in any C*-algebra, but
they are obtained from $U_G$
in the generalized Schr\"odinger representation.

We combine these C*-algebras into the kinematic field algebra, which is
\[
{\mathfrak A}_{\Lambda} :=
{\mathfrak F}_{\Lambda} \otimes \bigotimes_{\ell\in\Lambda^1}\big(C(G)\rtimes_\lambda G\big)
\]
which is is well--defined as $\Lambda^1$ is finite, and the cross--norms are unique as all
algebras in the entries are nuclear. (If $\Lambda^1$ is infinite, the tensor product
${\mathop{\bigotimes}\limits_{\ell\in\Lambda^1}\big(C(G)\rtimes_\lambda G\big)}$ is undefined, as
$C(G)\rtimes_\lambda G$ is nonunital).
Moreover, since $C(G)\rtimes_\lambda G\cong\cl K.\big(L^2(G)\big)$ and
$\cl K.(\cl H._1)\otimes\cl K.(\cl H._2)\cong\cl K.(\cl H._1\otimes\cl H._2),$ it follows
that
\[
\mathop{\bigotimes}\limits_{\ell\in\Lambda^1}\big(C(G)\rtimes_\lambda G\big)\cong\cl K.\big(
\mathop{\otimes}\limits_{\ell\in\Lambda^1}L^2(G)\big)\cong\cl K.(\cl L.)
\]
 as $\Lambda^1$ is finite, where $\cl L.$ is a generic infinite dimensional separable Hilbert space.
So
\[
{\mathfrak A}_{\Lambda} ={\mathfrak F}_{\Lambda} \otimes\mathop{\bigotimes}\limits_{\ell\in\Lambda^1}\big(C(G)\rtimes_\lambda G\big)\cong
{\mathfrak F}_{\Lambda} \otimes\cl K.\big(\mathop{\otimes}\limits_{\ell\in\Lambda^1}L^2(G)\big)\cong\cl K.(\cl L.)
\]
as ${\mathfrak F}_{\Lambda}$ is a full matrix algebra. This shows that for a finite lattice
there will be only one irreducible representation, up to unitary equivalence. Also, ${\mathfrak A}_{\Lambda}$ is
simple, so all representations are faithful.

The algebra ${\mathfrak A}_{\Lambda}$
is faithfully and irreducibly represented on $\cl H.=\cl H._F\otimes\mathop{\bigotimes}\limits_{\ell\in\Lambda^1}L^2(G)$
by $\pi=\pi_F\otimes\big(\mathop{\bigotimes}\limits_{\ell\in\Lambda^1}\pi_{\ell}\big)$
where $\pi_\ell:C(G)\rtimes_\lambda G\to L^2(G)$ is the generalized Schr\"odinger representation for the
$\ell^{\rm th}$ entry. Then $\pi\big({\mathfrak A}_{\Lambda}\big)$ contains in its multiplier algebra the
operators $\hat{T}^{(\ell)}_{\Phi_{ij}(\ell)},\;\hat{U}^{(\ell)}_g$ for all $\ell\in \Lambda^1$.

To complete the picture, we also need to define the action of the local gauge group on ${\mathfrak A}_{\Lambda}$.
Recall that in $\pi$ it is given by $\zeta\to{\rm Ad}(\hat{W}_\zeta)$, and this clearly preserves
$\pi\big({\mathfrak A}_{\Lambda}\big)=\cl K.(\cl H.)$ and defines a strongly continuous action $\alpha$
of $\gauc \Lambda$ on $\pi\big({\mathfrak A}_{\Lambda}\big)$ (hence on ${\mathfrak A}_{\Lambda} $)
 as $\zeta\to\hat{W}_\zeta$ is strong operator continuous.
By construction ${(\pi,\hat{W})}$ is a covariant representation for the
C*-dynamical system given by  $\alpha:\gauc \Lambda \to\aut{\mathfrak A}_{\Lambda}$.
As $\gauc \Lambda=\prod\limits_{x\in \Lambda^0}G$ is  compact, we can construct the crossed product
${\mathfrak A}_{\Lambda}\rtimes_\alpha \gauc \Lambda$ which has as representation space all covariant
representations of $\alpha:\gauc \Lambda \to\aut{\mathfrak A}_{\Lambda}$. As it is convenient to have an
identity in the algebra, our full field algebra for the system will be taken to be:
\[
\al F._e:=({\mathfrak A}_{\Lambda}\oplus\C)\rtimes_\alpha (\gauc \Lambda)
\]
where ${\mathfrak A}_{\Lambda}\oplus\C$ denotes ${\mathfrak A}_{\Lambda}$ with an identity adjoined.
This has a unique faithful representation on $\cH$ corresponding to the covariant representation
${(\pi,\hat{W})}$.

We consider two types of gauge invariant observables for lattice QCD which appeared in the literature (cf.~\cite{KS}).
\begin{itemize}
\item[(1)] We start with gauge invariant variables of pure gauge type, and consider the well-known Wilson loops cf.~\cite{Wilson}.
To construct a Wilson loop, we choose an oriented loop $L=\{\ell_1,\ell_2,\ldots,
 \ell_m\}\subset\Lambda^1$, $\ell_j=(x_j,y_j)$, such that  $y_j=x_{j+1}$ for $j=1,\ldots,{m-1}$
 and $y_m=x_1$. Let $G_k=G$ be the configuration space of $\ell_k$. Denoting the components of a gauge potential $\Phi$
 by $\Phi_{ij}(\ell_k)\in C(G_k)$ as in equation \eqref{Def-QuConn}, the matrix components of the quantum connection
at $\ell_k$ are given by $T_{\Phi_{ij}(\ell_k)}$.
To construct the gauge invariant observable associated with the loop, define (summing over repeated indices):
 \begin{eqnarray*}
 W(L)&:=&\Phi_{i_1i_2}(\ell_1)(g_1)\,\Phi_{i_2i_3}(\ell_3)(g_2)\cdots \Phi_{i_{m-1}i_1}(\ell_m)(g_m)\\[1mm]
 &=&(e_{i_1},g_1g_2\cdots g_me_{i_1})= {\rm Tr}(g_1g_2\cdots g_m)\,.
  \end{eqnarray*}
  (Note that to perform the product we need to fix identifications of $G_i$ with $G$.)
This defines a gauge invariant element $W(L)\in C(G_1)\otimes\cdots\otimes C(G_m)={C(G_1\times\cdots G_m)}$.
To see the gauge invariance, just note that
\[
\big(\zeta\cdot\Phi)(\ell)\big(\zeta\cdot\Phi)(\ell')=
\zeta(x_\ell)\,\Phi(\ell)\,\zeta(y_\ell)^{-1}\zeta(x_{\ell'})\,\Phi(\ell')\,\zeta(y_{\ell'})^{-1}
=\zeta(x_\ell)\,\Phi(\ell)\,\Phi(\ell')\,\zeta(y_{\ell'})^{-1}
\]
if $y_\ell=x_{\ell'}$ (i.e. $\ell'$ follows $\ell$), and use the trace property
for $W(L)$.

Wilson loops of particular importance are those where the paths are plaquettes,
i.e. $L={(\ell_1,\ell_2,\ell_3,\ell_4)}\in\Lambda^2$ as such $W(L)$ occur in the lattice Hamiltonian.
As remarked above, as $C(G_j)\subset M({C(G_j)\rtimes_\lambda G_j})$, it is not actually contained in $\cl L._{\ell_j}$.
We embed $C(G_1)\otimes\cdots\otimes C(G_m)$ (hence $W(L)$)
 in $M({\mathfrak A}_{\Lambda})$ by letting it act as the identity in entries not corresponding to
 $\{\ell_1,\ell_2,\ell_3,\ell_4\}.$

\item[(2)]  Another method of constructing gauge invariant observables, is
  by Fermi bilinears connected with a Wilson line (cf.~\cite{KS}).
  Consider a path  $C=\{\ell_1,\ell_2,\ldots,
 \ell_m\}\subset\Lambda^1$, $\ell_j=(x_j,y_j)$, such that  $y_j=x_{j+1}$ for $j=1,\ldots,m-1$.
 We take notation as above, so
 $G_k=G$ is the configuration space of $\ell_k,$ and
$\Phi_{ij}(\ell_k)(g_k):=(e_i,g_ke_j)$, $g_k\in G_k$.
To construct a gauge invariant observable associated with the path, consider (with summation convention):
 \begin{eqnarray*}
 Q(C)&:=& \psi^*_{i_1}(x_1)\,\Phi_{i_1i_2}(\ell_1)\,\Phi_{i_2i_3}(\ell_3)\cdots \Phi_{i_{m-1}i_m}(\ell_m)\, \psi_{ i_m}(y_m)\\[1mm]
&\in& {\mathfrak F}_{S}\otimes C(G_1)\otimes\cdots\otimes C(G_m)
  \end{eqnarray*}
where $S\subseteq\Lambda^0$ contains the path and we assume $\Cn=\C^k$
(otherwise $\Cn=\C^k\times{\bf W}$ and there are more indices). Then $ Q(C)$ is gauge invariant.
As above,
 we embed ${\mathfrak F}_{S}\otimes C(G_1)\otimes\cdots\otimes C(G_m)$
 in $M({\mathfrak A}_{\Lambda})$ in the natural way.
\end{itemize}

In the representation $\pi=\pi_F\otimes\big(\mathop{\bigotimes}\limits_{\ell\in\Lambda^1}\pi_{\ell}\big)$
 it is also possible to build unbounded observables
as gauge invariant operators.
For example we can  build gauge invariant combinations of the gluonic and the colour electric field
generators, and in the finite lattice context,
such operators were analyzed in \cite{KR1,JKR}.
As an example of such a gauge invariant operator, in the context of a finite lattice,
we state the Hamiltonian, where we disregard terms by which
$H$ has to be supplemented in order to avoid the
fermion doubling problem.
We first need to add spinor indices, hence take
$\Cn = \b W.\otimes\C^k$ where $\b W.\cong\C^4$ will be the spinor part on which the $\gamma$-matrices act.
 If $\{w_1,\ldots,w_4\}$ is an orthonormal basis of $\b W.$
and  $\{e_1,\ldots,e_k\}$ is an orthonormal basis of $\C^k$, then w.r.t. the orthonormal basis $\{w_j\otimes e_n\mid j=1,\ldots,4,\,
n =1,\ldots,k\}$ of $\Cn$, we obtain the  indices
\[
a(w_j\otimes e_n\cdot\delta_x)=:\psi_{jn}(x)
\]
for the quark field generators, where the subscript  $j$ is the spinor index, and the $n$ is the gauge index.
Then the Hamiltonian is
\begin{eqnarray}
\label{Hamiltonian}
  H & = & \tfrac{a}{2} \sum_{\ell \in \Lambda^1}
  E_{ij}(\ell) E_{ji}(\ell)
  + \tfrac{1}{2 g^2 a}\sum_{p \in \Lambda^2}
 ( W (p) + W(p)^*) \nonumber \\
  & + & i\tfrac{a}{2} \sum_{\ell \in \Lambda^1}
  \bar\psi_{jn}(x_\ell)  \big[\underline\gamma\cdot(y_\ell-x_\ell)\big]_{ji}
  \Phi_{nm} (\ell)\psi_{im} (y_\ell)  + h.c.
  \nonumber \\
  & + & ma^3 \sum_{x \in \Lambda^0} \bar \psi_{jn} (x)  \psi_{jn}(x)\,,
\end{eqnarray}
where $a$ is the assumed lattice spacing; $W (p)$ is the Wilson loop operator for the
plaquette $p= (\ell_1, \ell_2 ,\ell_3, \ell_4)$; the vector
$y_\ell-x_\ell$ for a link $\ell=(x_\ell,y_\ell)$  is the vector of length $a$ pointing from $x_\ell$ to $y_\ell$
 and h.c. means the Hermitean conjugate. As usual for spinors, $ \bar \psi_{jn} (x)=\psi_{in}(x)^*(\gamma_0)_{ij}$,
 and we use the standard gamma-matrices.
 We have omitted the flavour indices.
The summands occurring in \eqref{Hamiltonian} are either Laplacians, Wilson loop operators
or Fermi bilinears, hence they are all
 gauge invariant,  hence are
 observables, some unbounded.

The above  Hamiltonian suffers from the well-known fermion
doubling problem (cf. \cite{FL01});- the latter can be cured by passing e.g. to Wilson
fermions \cite{Wilson2}. This modification does  not affect the arguments below,
hence we focus our analysis on  the naive Hamiltonian given by (\ref{Hamiltonian}).
Below in Sect.~\ref{LDA} we will consider the dynamics produced by this Hamiltonian.


\subsection{The Fermion algebra for an infinite lattice.}
\label{FAlg}

It is unproblematic to specify the Fermion field on an infinite lattice $\Lambda=(\Lambda^0, \Lambda^1, \Lambda^2)$:\\[4mm]
\noindent {\bf Assumption:}
{\sl Assume the quantum matter field algebra on $\Lambda$ is:
\begin{equation}
\label{fermifieldalgebra}
{\mathfrak F}_{\Lambda} :=  \CAR \ell^2(\Lambda^0,\Cn). =C^*\big(\mathop{\bigcup}_{x\in\Lambda^0}{\mathfrak F}_x\big)
\end{equation}
where ${\mathfrak F}_x:=\CAR V_x.$ and $V_x:=\{f\in\ell^2(\Lambda^0,\Cn)\,\mid\, f(y)=0\;\;\hbox{if}\;\; y\not=x\}\cong \Cn.$
We interpret ${\mathfrak F}_x\cong\CAR \Cn.$ as the field algebra for a fermion at $x.$
We denote the generating elements of $\CAR \ell^2(\Lambda^0,\Cn).$ by $a(f),$ $f\in\ell^2(\Lambda^0,\Cn),$ and these
satisfy the usual CAR--relations:
\begin{equation}
  \label{eq:car}
\{a(f), a(g)^*\}  = \la f, g \ra \1
\quad \hbox{ and } \quad  \{a(f),a(g)\} = 0
\quad \mbox{ for } \quad f, g \in \ell^2(\Lambda^0,\Cn)
\end{equation}
where  $\{A,B\} := AB + BA$.}\\[4mm]
Note that the odd parts of ${\mathfrak F}_x$ and ${\mathfrak F}_y$
w.r.t. the fields $a(f)$ anticommute if $x\not=y.$ Moreover, as $\Lambda^0$ is infinite,
${\mathfrak F}_{\Lambda}$ has inequivalent irreducible representations.

We have the following inductive limit structure.
Let $\cl S.$ be a directed set of finite
connected subgraphs $S\subset(\Lambda^0, \Lambda^1)$,
such that $\bigcup\limits_{S\in\cl S.}S=(\Lambda^0, \Lambda^1)$,
 where the partial ordering is set inclusion.
Note that $S_1\subseteq S_2$ implies $\Lambda_{S_1}^i\subseteq\Lambda_{S_2}^i$
and $\bigcup\limits_{S\in\cl S.}\Lambda_S^i=\Lambda^i.$
Define
${\mathfrak F}_{S}:=C^*\big(\mathop{\cup}\limits_{x\in\Lambda_S^0}{\mathfrak F}_x\big)\subset{\mathfrak F}_{\Lambda}$
and then ${\mathfrak F}_{\Lambda}=\ilim{\mathfrak F}_{S}$ is an inductive limit w.r.t. the partial ordering
in $\cl S..$

This  defines the quantum matter fields on the lattice sites, and to obtain correspondence with the physics notation,
we choose an appropriate orthonormal basis in $\Cn$ and proceed as before for a finite lattice.


\subsection{A maximal C*-algebra  for dynamics construction.}
\label{GFA}


In this  section we wish to define a concrete C*-algebra which is ``maximally large'' in the sense that
 it contains all the true local C*-algebras, as well as the (bounded) terms of the local Hamiltonians.
 In the next section we will then define the dynamics automorphisms on it, and finally
we will then take our new field C*-algebra to be the smallest subalgebra
 which contains  all the true local C*-algebras, and is preserved w.r.t. the dynamics.
 Thus, the dynamics itself will create the field algebra for the system.
 The gauge transformations will be added later.

Recall from above that for every link $\ell$ we have a generalised Weyl  algebra ${C(G)\rtimes_\lambda G}
\cong\cl K.\big(L^2(G)\big)$, for which
we have a generalized Schr\"odinger representation
 $\pi_0:C(G)\rtimes_\lambda G\to\cl B.\big(L^2(G)\big)$
such that
 $\pi_0\big(C(G)\rtimes_\lambda G\big)=\cl K.\big(L^2(G)\big)$.
  Explicitly,
it is given by
  $\pi_0(\varphi\cdot f)=\pi_1(\varphi)\pi_2( f)$ for $\varphi\in L^1(G)$ and  $f\in C(G)$ where
 \[
(\pi_1(\varphi)\psi)(g):=\int\varphi(h)\psi(h^{-1}g)\,{\rm d}h \qquad\hbox{and}\qquad
 (\pi_2(f)\psi)(g):=f(g)\psi(g)
 \]
 for all $\psi\in L^2(G)$. Note that by irreducibility, the constant vector $\psi_0=1$ is cyclic
 and normalized (assuming normalized Haar measure on $G$).

 We start by taking an infinite product of generalized Schr\"odinger representations,
 one for each $\ell\in\Lambda^1$, w.r.t. the reference sequence ${(\psi_0,\psi_0,\ldots)}$
where $\psi_0=1$ is the constant vector (cf.\cite{vN}). Thus the
space $\al H._\infty$  is the completion of the pre--Hilbert space spanned by finite combinations of
elementary tensors of the type
\[
\varphi_1\otimes\cdots\otimes \varphi_k\otimes\psi_0\otimes\psi_0\otimes\cdots,\quad\varphi_i\in \al H._i= L^2(G),\;
k\in\N
\]
w.r.t. the pre--inner product given by
\[
\left(\varphi_1\otimes\cdots\otimes \varphi_k\otimes\psi_0\otimes\psi_0\otimes\cdots,\;
\varphi'_1\otimes\cdots\otimes \varphi'_k\otimes\psi_0\otimes\psi_0\otimes\cdots\right)_\infty
:=\prod_{i=1}^k(\varphi_i,\varphi'_i)\,,
\]
and we assume the usual entry-wise tensorial linear operations for the elementary tensors.
Denote the reference vector by $\psi_0^\infty:=\psi_0\otimes\psi_0\otimes\cdots$.

Fix a  finite nonempty connected subgraph $S$ of $(\Lambda^0, \Lambda^1)$, and let
 \[
 \Lambda_S:=(\Lambda^0_S, \Lambda^1_S, \Lambda^2_S).
 \]
  Then
 $\al L._S:={\mathop{\bigotimes}\limits_{\ell\in\Lambda^1_S}\big(C(G)\rtimes_\lambda G\big)}$ acts on  $\al H._\infty$
  as a product representation of $\pi_0$ where each factor $\al L._\ell:=C(G)\rtimes_\lambda G$
 acts on the factor of $\al H._\infty$ corresponding to $\ell$. In fact if $[\cdot]$ denotes closed span, then
 \[
[\al L._S\psi_0^\infty]=\Big(\mathop{\bigotimes}\limits_{\ell\in\Lambda^1_S}\al H._\ell\Big)\otimes
 \mathop{\bigotimes}\limits_{\ell\not\in\Lambda^1_S}\psi_0\subset\al H._\infty.
 \]
 Thus all $\al L._S$ are faithfully imbedded in $\al B.(\al H._\infty)$, and if $S$ and $S'$ are disjoint, $\al L._S$
 and $\al L._{S'}$ commute.

 Now consider the Fock representation $\pi_{\rm Fock}:{\mathfrak F}_{\Lambda}\to\al B.(\al H._{\rm Fock})$
 of the CAR--algebra with vacuum vector $\Omega$. The Hilbert space on which we define our infinite lattice model is
 \[
 \al H.:=\al H._{\rm Fock}\otimes\al H._\infty.
 \]
 Then by
 ${\mathfrak F}_{S}\subset{\mathfrak F}_{\Lambda}$, we also have a product representation of the local field algebras
 ${\mathfrak A}_S:={\mathfrak F}_{S}\otimes\mathop{\bigotimes}\limits_{\ell\in\Lambda^1_S}\cl L._\ell$
 on $\al H.$. If $S$ and $S'$ are disjoint, then ${\mathfrak A}_S$ and  ${\mathfrak A}_{S'}$
 will graded--commute w.r.t. the Fermion grading.
If we have containment i.e., $R\subset S$, then ${\mathfrak F}_{R}\subset{\mathfrak F}_{S}$, but we have
$\mathop{\bigotimes}\limits_{\ell\in\Lambda^1_R}\cl L._\ell\not\subset
 \mathop{\bigotimes}\limits_{\ell\in\Lambda^1_S}\cl L._\ell$ because
 $\al K.(\al H._1)\otimes\un\not\subset\al K.(\al H._1\otimes\al H._2)$
 if $\al H._2$ is infinite dimensional. However w.r.t. the natural operator product
 we have  ${\mathfrak A}_R\cdot{\mathfrak A}_S={\mathfrak A}_S$, hence
 ${\mathfrak A}_R\subset M({\mathfrak A}_S)$, as the action of  ${\mathfrak A}_R$ on ${\mathfrak A}_S$
is nondegenerate.

 Note that the Fermion grading also produces a grading unitary
 $U_F$ on $\al H._{\rm Fock}$ as the second quantization of -1 acting on $\ell^2(\Lambda^0,\Cn)$. This grading
coincides with the even-odd grading for  $n\hbox{-particle}$  vectors in $\al H._{\rm Fock}$.
 Naturally  $U_F$ extends to
  $\al H.:=\al H._{\rm Fock}\otimes\al H._\infty$ as $U_F\otimes\un$, which extends by conjugation
   the Fermion grading to all of $\cl B.(\al H.)$.

 The local graded commutation properties above are crucial for our construction of local dynamics, and
 this property will be preserved
 in our definition of a maximal C*-algebra on which we will construct the dynamics.
 For each finite connected subgraph $S$ of $(\Lambda^0, \Lambda^1)$, we define
\[
\al H._S:= [{\mathfrak F}_{S}\Omega]\otimes[\al L._S\psi_0^\infty],\qquad\cl B._S:= \cl B.(\al H._S).
\]
To see how to imbed $ \cl B.(\al H._S)\subset \cl B.(\al H.)$,
note that ${\mathfrak F}_{S}$ is finite dimensional, hence so is $[{\mathfrak F}_{S}\Omega]$
and as the
restriction of $\pi_{\rm Fock}({\mathfrak F}_{S})$ to $[{\mathfrak F}_{S}\Omega]$ is just the Fock representation of
${\mathfrak F}_{S}$, we obtain from irreducibility that $\cl B.([{\mathfrak F}_{S}\Omega])= {\mathfrak F}_{S}$ on
 $[{\mathfrak F}_{S}\Omega]$. Thus
\[
\cl B._S=\cl B.(\al H._S)=\cl B.( [{\mathfrak F}_{S}\Omega]\otimes[\al L._S\psi_0^\infty])
 =\cl B.([{\mathfrak F}_{S}\Omega])\otimes \cl B.([\al L._S\psi_0^\infty])
={\mathfrak F}_{S}\otimes\cl B.([\al L._S\psi_0^\infty])
\]
where the third equality follows from \cite[Example~11.1.6]{KR83}.
Let ${\mathfrak F}_{S}\subset{\mathfrak F}_{\Lambda}$ act on $\al H._{\rm Fock}$ as part of the Fock representation of
${\mathfrak F}_{\Lambda}$. As
\[
[\al L._S\psi_0^\infty]=\Big(\mathop{\bigotimes}\limits_{\ell\in\Lambda^1_S}\al H._\ell\Big)\otimes
 \mathop{\bigotimes}\limits_{\ell\not\in\Lambda^1_S}\psi_0\subset\al H._\infty,
\]
we can extend $\cl B.([\al L._S\psi_0^\infty])$ to $\al H._\infty$ by letting its elements act as the identity
on those factors of $\al H._\infty$ corresponding to $\ell\not\in\Lambda^1_S$, i.e.
\begin{equation}
\label{BSisFB}
\cl B._S
=\pi_{\rm Fock}({\mathfrak F}_{S})\otimes\cl B.\Big(\mathop{\bigotimes}\limits_{\ell\in\Lambda^1_S}\al H._\ell\Big)
\otimes\mathop{\bigotimes}\limits_{\ell\not\in\Lambda^1_S}\un.
\end{equation}
 Thus we have obtained the embedding
$ \cl B.(\al H._S)\subset \cl B.(\al H.)$, and we have containments $ \cl B.(\al H._S)\subseteq \cl B.(\al H._T)$
if $S\subseteq T$.
Note that the restriction of this embedded copy of $ \cl B.(\al H._S)$ to $\al H._S\subset\al H.$
gives a faithful representation, but it is nonzero
outside $\al H._S$ as it contains the identity.
Now we can define:
\[
\al A._{\rm max}:=\ilim \cl B._S = C^*\Big(\bigcup_{S\in\al S.}  \cl B.(\al H._S)   \Big)
\]
 where in the last inductive limit and union, $S$ ranges over the directed set $\al S.$  of all
 finite connected subgraphs of $(\Lambda^0, \Lambda^1)$.
 Here and below, we will assume that $\al S.$ does not contain the empty set, so we can ignore this trivial special case.

 Note that if we use the grading unitary above to extend the Fermion grading to all of $\cl B.(\al H.)$,
 then if $S$ and $S'$ are disjoint, then $\cl B.(\al H._S)\supset{\mathfrak A}_S$ and  $\cl B.(\al H._{S'})
 \supset{\mathfrak A}_{S'}$ will graded--commute.
  Note that a similar inductive limit
cannot be done for the local algebras ${\mathfrak A}_S$ as it is NOT true that
 $S\subset S'$ implies ${\mathfrak A}_S\subset{\mathfrak A}_{S'}$.
 An operator $A\in\cl B.(\al H.)$ will be said to have {\it support in} $S$ if $S$ is the smallest
 connected graph for which $A\in\cl B.(\al H._S)=\cl B._S$.

As $\al A._{\rm max}$ contains all the local algebras ${\mathfrak A}_S$, we aim to define
the full dynamics on $\al A._{\rm max}$, and then generate the new field algebra in $\al A._{\rm max}$
from the orbits of the local ones. At the end of the next section we will show that
$\al A._{\rm max}$  is contained in the multiplier algebra of the kinematics field algebra
we constructed previously in \cite{GrRu}.

\section{Dynamics}
\label{Dynamics}

Next, we want to define the dynamics on $\al A._{\rm max}:=\ilim \cl B._S $ corresponding to the heuristic
Hamiltonian given by
\begin{eqnarray}
\label{HamiltonianInft}
   H & = & \tfrac{a}{2} \sum_{\ell \in \Lambda^1}
  E_{ij}(\ell) E_{ji}(\ell)
  + \tfrac{1}{2 g^2 a}\sum_{p \in \Lambda^2}
 ( W (p) + W(p)^*) \nonumber \\
  & + & i\tfrac{a}{2} \sum_{\ell \in \Lambda^1}
  \bar\psi_{jn}(x_\ell)  \big[\underline\gamma\cdot(y_\ell-x_\ell)\big]_{ji}
  \Phi_{nm} (\ell)\psi_{im} (y_\ell)  + h.c.
  \nonumber \\
  & + & ma^3 \sum_{x \in \Lambda^0} \bar \psi_{jn} (x)  \psi_{jn}(x)\,,
\end{eqnarray}
where the difference with (\ref{Hamiltonian}) is that now the sums are over an infinite lattice, so are
not yet properly defined. The unbounded operators in the first summand are defined on the appropriate factor
of $\al H._\infty$, and all calculations will be done concretely, i.e. in terms of operators on
$\al H.:=\al H._{\rm Fock}\otimes\al H._\infty$, and we will not explicitly indicate the
representations $\pi_0$ and $\pi_{\rm Fock}$.  Otherwise, notation
 is as before.

The summands occurring in \eqref{HamiltonianInft} are all gauge invariant (locally) and hence are
 observables, some unbounded.
 All summands also have Fermion degree zero, hence for these summands graded local commutativity becomes just
ordinary local commutativity
 with elements of the algebra $\al A._{\rm max}:=\ilim \cl B._S $.

 Below we will follow the familiar technique for defining the dynamics of lattice systems by first defining it for
 each $S\in\cl S.$, then proving that these ``local automorphism groups" have a pointwise norm limit which defines
 a dynamics automorphism for the full algebra (cf. \cite{BR2}).

\subsection{The local dynamics automorphism groups.}
\label{LDA}

Now for a fixed $\ell$, in the representation $\pi_0$ on $L^2(G)$ the operator  $E_{ij}(\ell) E_{ji}(\ell)$ (summation convention)
is just the group Laplacian, hence it is defined on the domain
 $C^\infty(G)\subset L^2(G)$, and it is essentially selfadjoint as
it produces a positive quadratic
form. These are the only unbounded terms in $H$. Given $S\in\cl S.$ we define the local Hamiltonian $H_S$
by summing only over the restricted lattice $\Lambda_{S}$, thus:
\begin{eqnarray*}
H_S&=& H_S^{\rm loc}+H_S^{\rm int}\qquad\hbox{on}\qquad \cl H._{\rm Fock}\otimes\cl D._S
\qquad\hbox{where} \\[2mm]
 H_S^{\rm loc} & := & \tfrac{a}{2} \sum_{\ell \in \Lambda^1_S}
  E_{ij}(\ell) E_{ji}(\ell)+ma^3 \sum_{x \in \Lambda^0_S} \bar \psi_i (x)  \psi_i(x)
  \qquad\hbox{on}\qquad \cl H._{\rm Fock}\otimes\cl D._S \\[1mm]
 H_S^{\rm int} & := & \tfrac{1}{2 g^2 a}\sum_{p \in \Lambda^2_S}
 ( W (p) + W(p)^*) +i\tfrac{a}{2} \sum_{\ell \in \Lambda^1_S}
  \bar\psi_{jn}(x_\ell)  \big[\underline\gamma\cdot(y_\ell-x_\ell)\big]_{ji}
  \Phi_{nm} (\ell)\psi_{im} (y_\ell) + h.c.
\end{eqnarray*}
where $\cl D._S\subset\bigotimes\limits_{\ell\in\Lambda^1}L^2(G)$ is the span of those elementary tensors
such that if $\ell\in\Lambda^1_S$, then in the $\ell\hbox{-th}$ entry it takes its value in $C^\infty(G)$
but otherwise it is unrestricted. One may write this as
$\cl D._S=\bigotimes\limits_{\ell\in\Lambda^1_S}C^\infty(G)\otimes\bigotimes\limits_{\ell'\not\in\Lambda^1_S}L^2(G)$
(infinite tensor products are w.r.t. the reference vector $\psi_0^\infty$).
Note that $H_S^{\rm loc}$ is an unbounded essentially selfadjoint (positive) operator which  affects the individual
sites and links independently (so produces free time evolution), and that the interaction term $H_S^{\rm int}$ is bounded.
In fact the free time evolution is just a tensor product of the individual free time evolutions:
\[
U^{\rm loc}_S(t):=\exp(it\bar{H}_S^{\rm loc})=U^{\rm CAR}_S(t)\otimes
\bigotimes_{\ell\in\Lambda^1_S}U_\ell(t)\otimes\bigotimes\limits_{\ell'\not\in\Lambda^1_S}\1
\]
where  $U^{\rm CAR}_S(t)=\exp\big(itma^3 \sum\limits_{x \in \Lambda^0_S} \bar \psi_i (x)  \psi_i(x)\big)\in{\mathfrak F}_{S}$
and $U_\ell(t):=\exp(it\overline{E_{ij}(\ell) E_{ji}(\ell)})$
(the overline notation $\bar{H}_S^{\rm loc}$ and $\overline{E_{ij}(\ell) E_{ji}(\ell)}$ indicates  the closure
of the essentially selfadjoint operators).
 The local free time evolutions
$\alpha^{\rm loc}_S(t):={\rm Ad}(U^{\rm loc}_S(t))$ will preserve each $\cl B._{S'}, $ $S'\in\al S.$, hence preserves
$\al A._{\rm max}=\ilim \cl B._S $ because it acts componentwise. However $\alpha^{\rm loc}$ is not strongly continuous
i.e. pointwise norm continuous, as $H_S^{\rm loc}$ is unbounded and  $\cl B._S=\cl B.(\al H._S)$ (cf. Prop.~5.10 in \cite{GrN14}).

 As
$U^{\rm CAR}_S(t)\in{\mathfrak F}_{S}\subset{\mathfrak F}_{\Lambda},$ we have that $\alpha^{\rm CAR}_S(t):={\rm Ad}(U^{\rm CAR}_S(t))$
clearly preserves both ${\mathfrak F}_{S}$ and ${\mathfrak F}_{\Lambda}.$ Moreover $t\mapsto\alpha^{\rm CAR}_S(t)$ is uniformly norm continuous
as  the generator $ma^3 \sum\limits_{x \in \Lambda^0_S} \bar \psi_i (x)  \psi_i(x)$ is bounded.

Observe that $\alpha^{\rm loc}_S(t)$ satisfies the compatibility condition that if $R\subset S$, then $\alpha^{\rm loc}_S(t)$ preserves the subalgebra
$\al B._R\subset\al B._S$ and its restriction on it coincides with $\alpha^{\rm loc}_R(t)$, as
is clear from the tensor product construction.
We have that
${\rm Ad}(U^{\rm loc}_S(t))$ defines a  one parameter group  $t\mapsto\alpha^{\rm loc}_S(t)\in{\rm Aut}(\al A._{\rm max})$
which preserves  $\al B._S$, and is the identity on any $\al B._R$ where $R\cap S=\emptyset$.

Next, note that as the (bounded) interaction Hamiltonian $H_S^{\rm int}\in \al B._S,$ its commutator produces a derivation on
$\al A._{\rm max}$, which preserves $\al B._S$ i.e.
$ [H_S^{\rm int},\al B._S]\subseteq \al B._S$.

To be more precise, consider the individual terms in the finite sums comprising $ H_S^{\rm int}$.
 Recall that for $p={(\ell_1,\ell_2,\ell_3,\ell_4)}\in\Lambda^2$, we have
\begin{eqnarray*}
W(p)&\in& C(G_{\ell_1})\otimes\cdots\otimes C(G_{\ell_4})={C(G_{\ell_1}\times\cdots G_{\ell_4})}\subset M(
{\cal L}_{S}) \subseteq  \al B._S  \qquad
\hbox{if}\qquad p\subset S\\[1mm]
\hbox{and}\;
&&   \bar\psi_{jn}(x_\ell)  \big[\underline\gamma\cdot(y_\ell-x_\ell)\big]_{ji}
  \Phi_{nm} (\ell)\psi_{im} (y_\ell)\in {\mathfrak F}_{S}\otimes C(G_\ell)\subset \al B._S
 \end{eqnarray*}
where $\ell=(x,y)\subset S$.
Thus the commutator with $ H_S^{\rm int}$ defines a bounded derivation on
$\al A._{\rm max}$, which preserves $\al B._S$.

For each $S\in\cl S.$ we have a
local time evolution $\alpha^S:\R\to{\rm Aut}\big(\al A._{\rm max}\big)$ which preserves
$\pi(\al B._S)$ and acts trivially on $\al A._{\rm max}$ outside of it, given by
\[
\alpha^S_t:={\rm Ad}(U_S(t))\quad\hbox{and}\quad
U_S(t):=\exp(itH_S).
\]
  Due to the interaction terms, we do not expect
that  $R\subset S$ implies that $\alpha^S$ will coincide with $\alpha^R$ on  $\al B._R\subset\al B._S$.
By construction, $(\pi,U_S)$ is a covariant irreducible representation for $\alpha_S:\R\to{\rm Aut}(\al A._{\rm max})$.
The  generator $H_S= H_S^{\rm loc}+H_S^{\rm int}$ of its implementing unitary group $U_S$ has a positive unbounded part
$H_S^{\rm loc}$ and a bounded interaction part $H_S^{\rm int}$, hence it is bounded from below.
As $H_S$ is unbounded,  $\alpha^S$ is not strongly continuous on $\al B._S$. However, on the local algebras
 ${\mathfrak A}_S\subset\al B._S$ we have:
\begin{Lemma}
\label{localauto}
Let $R,\,S\in\al S.$ with $R\subseteq S$.
For each $A\in{\mathfrak A}_S$ we have that $\; t\mapsto \alpha^R_t(A)\;$ is continuous, i.e. the restriction
of $\alpha^R$ to ${\mathfrak A}_S$ is strongly continuous.
\end{Lemma}
\begin{beweis}
Recall that
${\mathfrak A}_S:={\mathfrak F}_{S}\otimes\mathop{\bigotimes}\limits_{\ell\in\Lambda^1_S}\cl L._\ell$
acts faithfully  on
\[
\al H._S:= [{\mathfrak F}_{S}\Omega]\otimes[\al L._S\psi_0^\infty]\subset\cl H.\quad\hbox{where}\quad
[\al L._S\psi_0^\infty]=\Big(\mathop{\bigotimes}\limits_{\ell\in\Lambda^1_S}\al H._\ell\Big)\otimes
 \mathop{\bigotimes}\limits_{\ell\not\in\Lambda^1_S}\psi_0\subset\al H._\infty.
 \]
As $S$ is finite, $[{\mathfrak F}_{S}\Omega]$ is finite dimensional, and so by $\al L._S
 =\al K.\Big(\mathop{\bigotimes}\limits_{\ell\in\Lambda^1_S}\al H._\ell\Big)$ we conclude that
${\mathfrak A}_S\restriction\al H._S\subseteq\al K.(\al H._S)$. Now  $\alpha^R_t={\rm Ad}(U_R(t))$
where $U_R(t)=\exp(itH_R)$, and $U_R(t)$ preserves $\al H._S$ and is the identity on any $\al H._{S'}$
where $S'$ is disjoint from $S$. Thus for $A\in{\mathfrak A}_S$, $\alpha^R_t(A)$ becomes just the conjugation of a
compact operator by a strong operator continuous one parameter unitary group, and this is well known to be norm continuous
in the parameter
(the strict topology on $\al B.(\al H.)=M(\al K.(\al H.))$ is the $\sigma\hbox{--strong}$ *--topology).
\end{beweis}
The analogous converse statement  will be proven below in the proof of Theorem~\ref{GlobDynCont}, i.e.
that if  $R\subset S$, then $\; t\mapsto \alpha^S_t(A)\;$ is continuous for $A\in{\mathfrak A}_R$.
As ${\mathfrak A}_R\not\subset{\mathfrak A}_S$ and $\alpha^S_t$ need not preserve ${\mathfrak A}_R$,
this is not obvious.

\subsection{Existence of the global dynamics automorphism group.}
\label{DMFA}

We want to apply arguments of Nachtergaele and Sims  in \cite{NaSi} to establish the existence of the full
infinite lattice dynamics on $\al A._{\rm max}$.
Here we will supply all the necessary details to adapt their argument to our situation.

Return to the faithful representation
$\pi=\pi_{\rm Fock}\otimes\pi_\infty$, on $\cl H.=\cl H._{\rm Fock}\otimes\cl H._\infty$ where
$\cl H._\infty=\bigotimes\limits_{\ell\in\Lambda^1}L^2(G_\ell)$ is defined w.r.t. the reference vector
 $\psi_0\otimes\psi_0\otimes\cdots$.
 All calculations below will be done concretely in this representation, but to limit notation
 it will not usually be indicated.
 As before, we fix  the directed set $\cl S.$ of  finite connected subgraphs of $(\Lambda^0,\Lambda^1)$,
 where the partial ordering is set inclusion.
 Note that for each link $\ell\in\Lambda^1$, we can identify a graph in $\cl S.$ (also denoted by $\ell$)
 as the graph consisting of the endpoints $(x_\ell,y_\ell)$ for vertices, and the edge from $x_\ell$ to
 $y_\ell$. Likewise, we can identify any plaquette $p\in\Lambda^2$ with a graph in $\cl S.$.
 Then
 \[
 \al A._{\rm max}=\ilim \cl B._S = C^*\Big(\bigcup_{S\in\al S.}  \cl B.(\al H._S)   \Big)
 \qquad\hbox{and}\qquad
 {\mathfrak F}_{S}\otimes{\cal L}_{S}
 \subset \cl B.(\al H._S).
\]
Recall that for each $S\in\cl S.$ we have a
local time evolution $\alpha^S:\R\to{\rm Aut}\big(\al A._{\rm max}\big)$ which preserves
$\pi(\al B._S)$ and acts trivially on $\al A._{\rm max}$ outside of it, given by $\alpha^S_t={\rm Ad}(U_S(t))$.
We will show below that for each $A\in \al B._R,$ $R\subset S$
 that $\alpha^S_t(A)$ converges as ${S\nearrow\Z^3},$
where ${S\nearrow\Z^3}$ indicates that we take the  limit over increasing sequences in $\cl S.$ such that
 the union of the graphs in the sequence is the entire connected graph ${(\Lambda^0,\,\Lambda^1)}$ for the lattice.

We now want to apply arguments in \cite{NaSi} to establish the existence of an
infinite lattice dynamics on $ \al A._{\rm max}$.
We first revisit notation. Fix $S\in\cl S.$, then
\begin{eqnarray*}
H_S&=& H_S^{\rm loc}+H_S^{\rm int}\qquad\hbox{on}\qquad \cl H._{\rm Fock}\otimes\cl D._S
\qquad\hbox{where} \\[2mm]
 H_S^{\rm loc} & := & \sum_{\ell \in \Lambda^1_S}H_\ell+\sum_{x \in \Lambda^0_S}H_x\qquad\hbox{where}\\[2mm]
 H_\ell&:=&\tfrac{a}{2}
  E_{ij}(\ell) E_{ji}(\ell)\qquad\hbox{and}\qquad H_x:=ma^3  \bar \psi_i (x)  \psi_i(x)\\[2mm]
 H_S^{\rm int} & := & \sum_{p \in \Lambda^2_S}\widetilde{W}(p) +  \sum_{\ell \in \Lambda^1_S}B(\ell)
\qquad\hbox{where}\quad \widetilde{W}(p):= \tfrac{1}{2 g^2 a}( W (p) + W(p)^*)  \\[2mm]
\hbox{and}\qquad B(\ell)&:=&
i\tfrac{a}{2}
  \bar\psi_{jn}(x_\ell)  \big[\underline\gamma\cdot(y_\ell-x_\ell)\big]_{ji}
  \Phi_{nm} (\ell)\psi_{im} (y_\ell) + h.c.
\end{eqnarray*}
Clearly $H_S^{\rm int}\in \al B._S$ and $\sum_{x \in \Lambda^0_S}H_x\in \al B._S.$ The only terms of $H_S$ not in
the $ \al B._S$ are the unbounded $H_\ell$. Moreover the operator norm $\| \widetilde{W}(p)\|=:\|\widetilde{W}\|$ is independent of $p$
and  $\|B(\ell)\|=:\|B\|$ is independent of $\ell.$
Below we will frequently need the following notation:-
if $A\in\al B._S$ then
\[
A(t):=e^{itH_S^{\rm loc}}A\,e^{-itH_S^{\rm loc}}\in\al B._S\,.
\]
\begin{Theorem}
\label{GlobDynExist}
With notation as above, we have for all $A\in \al A._{\rm max}$ and $t\in\R$ that the norm limit
\[
\lim_{S\nearrow\Z^3}\alpha^S_t(A)=:\alpha_t(A)
\]
exists, and defines an automorphism group $t\mapsto\alpha_t\in{\rm Aut}(\al A._{\rm max})$.
Furthermore, for each $T>0$, the limit is uniform w.r.t. $t\in[-T,T]$.
\end{Theorem}
\noindent {\bf Proof:}
Fix $T>0,$ a nonempty $R\subset S$ and let $A\in\al B._R$. First we want to show that the
limit of $\alpha^S_t(A)$ as ${S\nearrow\Z^3}$ exists for $|t|<T$.
Fix a strictly increasing sequence
$\{S_n\}_{n\in\N}\subset\cl S.$ such that ${S_n\nearrow\Z^3}$ as $n\to\infty$.
The limit we obtain below will be independent of the choice of sequence
$\{S_n\}_{n\in\N}\subset\cl S.$ (given any two such sequences, each term of one sequence
will be contained in some term of the other one, which allows one to form a new increasing sequence
containing elements of both).

For the proof, we need to show
that $\{\alpha^{S_n}_t(A)\}_{n\in\N}$ is Cauchy. Now for $R\subseteq S_n$:
\[
\alpha^{S_n}_t(A)=\tau^{S_n}_t\big( e^{itH_{S_n}^{\rm loc}} A e^{-itH_{S_n}^{\rm loc}} \big)
=\tau^{S_n}_t\big( e^{itH_R^{\rm loc}} A e^{-itH_R^{\rm loc}} \big)=\tau^{S_n}_t\big( A(t) \big)
\]
where $\tau^{S}_t:={\rm Ad}(e^{itH_S}e^{-itH_S^{\rm loc}})$ so as
 $A(t)\in\al B._R$, it suffices to show that the sequence  $\{\tau^{S_n}_t(A)\}_{n\in\N}$ is Cauchy
for all $A\in\al B._R$.

Fix $S\in\cl S.$, consider the strong operator continuous map $U_S:\R\times\R\to\cl U.(\cl H.)$ given by
\[
U_S(t,s):=e^{itH_S^{\rm loc}}e^{i(s-t)H_S}e^{-isH_S^{\rm loc}}
\]
and note that $U_S(t,t)=\un$, $U_S(t,s)^*=U_S(s,t)$ and  $\tau^{S}_t(A)=U_S(0,t)AU_S(t,0)$.
As $H_S$ and $H_S^{\rm loc}$ differ by a bounded operator, they have the same domain $\cl D.$
and this domain is preserved by both unitary groups $e^{itH_S}$ and $e^{itH_S^{\rm loc}}$, hence by $U_S(t,s)$
and so for $\psi\in\cl D.$ we have
\[
\frac{d}{dt}U_S(t,s)\psi=ie^{itH_S^{\rm loc}}( H_S^{\rm loc}- H_S )e^{i(s-t)H_S}e^{-isH_S^{\rm loc}}\psi
=-iH_S^{\rm int}(t)U_S(t,s)\psi
\]
where  $H_S^{\rm int}(t):=e^{itH_S^{\rm loc}}H_S^{\rm int}e^{-itH_S^{\rm loc}}$. Likewise, for $\psi\in\cl D.$ we have
\[
\frac{d}{ds}U_S(t,s)\psi
=iU_S(t,s)H_S^{\rm int}(s)\psi.
\]
Now by Lemma~\ref{Lemma1} in the Appendix, we conclude that these relations hold on all of  $\cl H.$.
Let $R\subset S_n\subset S_m$ (hence $n<m$) then by the fundamental theorem of calculus
\[
\tau^{S_m}_t(A)-\tau^{S_n}_t(A)=\int_0^t\frac{d}{ds}\Big(U_{S_m}(0,s)U_{S_n}(s,t)A\,U_{S_n}(t,s)U_{S_m}(s,0)\Big)\,ds
\]
where the differential and integral is w.r.t. the strong operator topology.
The integrand is for any $\psi\in\cl H.$:
\begin{eqnarray*}
&&\!\!\!\!\!\!\!\!\!\!\!\!\!
\frac{d}{ds}\,U_{S_m}(0,s)U_{S_n}(s,t)A\,U_{S_n}(t,s)U_{S_m}(s,0)\psi\\[1mm]
&=&iU_{S_m}(0,s)\Big[(H_{S_m}^{\rm int}(s)-H_{S_n}^{\rm int}(s)),  U_{S_n}(s,t)A\,U_{S_n}(t,s)\Big]U_{S_m}(s,0)\psi\\[1mm]
&=&iU_{S_m}(0,s)e^{isH_{S_n}^{\rm loc}}\Big[{N}(s),  \alpha^{S_n}_{s-t}(A(t))\Big]e^{-isH_{S_n}^{\rm loc}}U_{S_m}(s,0)\psi
\qquad\hbox{where}\\[3mm]
{N}(s)&:=&e^{-isH_{S_n}^{\rm loc}}(H_{S_m}^{\rm int}(s)-H_{S_n}^{\rm int}(s))e^{isH_{S_n}^{\rm loc}}
=e^{isH_{S_m\backslash S_n}^{\rm loc}} H_{S_m}^{\rm int}e^{-isH_{S_m\backslash S_n}^{\rm loc}} -H_{S_n}^{\rm int}\\[1mm]
&=&{\rm Ad}\Big(e^{isH_{S_m\backslash S_n}^{\rm loc}}  \Big)\Big(\sum_{p \in \Lambda^2_{S_m}
\backslash \Lambda^2_{S_n}}\widetilde{W}(p) +  \sum_{\ell \in \Lambda^1_{S_m}\backslash \Lambda^1_{S_n}}B(\ell)\Big)\\[2mm]
&=&{\rm Ad}\Big(e^{isH_{S_m\backslash S_n}^{\rm loc}}  \Big)\Big(\sum_{q \in \Lambda^i_{S_m}
\backslash \Lambda^i_{S_n}}\Psi(q)\Big)
\qquad\qquad\hbox{and}\\[2mm]
 A(t)&=&  e^{itH_R^{\rm loc}} A e^{-itH_R^{\rm loc}}=e^{itH_{S_n}^{\rm loc}} A e^{-itH_{S_n}^{\rm loc}}
\end{eqnarray*}
where the last sum is over $\Lambda^i_{S_m}\backslash \Lambda^i_{S_n}:=\Lambda^1_{S_m}\backslash \Lambda^1_{S_n}
\cup\Lambda^2_{S_m}\backslash \Lambda^2_{S_n}$ and $\Psi(p):=\widetilde{W}(p)$ for plaquette $p$, and
$\Psi(\ell):=B(\ell)$ for a link $\ell$. Then $\|\Psi\|:=\max\{\|\tilde{W}\|,\|B\|\}\geq\|\Psi(q)\|$ for all $q$.

The support of ${N}(s)$ is contained in $S_m\backslash(S_n)_0$ which is defined as the set of
all the lattice points in $S_m$ (and links between them)
obtained from either links in $\Lambda^1_{S_m}\backslash \Lambda^1_{S_n}$ or plaquettes in
 $\Lambda^2_{S_m}\backslash \Lambda^2_{S_n}$.
Now ${N}(s)$ is a sum of terms with support in $q \in \Lambda^i_{S_m}
\backslash \Lambda^i_{S_n}$, and only those for which $q$ has a point in $S_n$ will have nonzero commutant with
$\al B._{S_n} $. Thus for $B\in \al B._{S_n}$, we have
\begin{eqnarray}
[N(s),B]&=& [\tilde{N}(s),B]\qquad\hbox{where}\nonumber\\[1mm]
\label{primesum}
\tilde{N}(s) & :=  &
{\rm Ad}\Big(e^{isH_{S_m\backslash S_n}^{\rm loc}}  \Big)\Big(\mathop{\mathord{\sum}'}_{q \in \Delta_{S_m}(
S_n)}\Psi(q)\Big)\qquad\hbox{and}\\[1mm]
\label{bdry}
\Delta_{T}(R)&:=&\{Z\subset T\,\big|\,Z\cap R\not=\emptyset\not=Z\cap(T\backslash R)\},
\end{eqnarray}
and the prime on the sum indicates that it is restricted by requiring $q$ to be a link or plaquette.

Thus, using Lemma~\ref{Lemma2} in the Appendix, and the fact that $\cl H.$ is separable, we get
\begin{eqnarray}
\big\|\tau^{S_m}_t(A)-\tau^{S_n}_t(A)\| &\leq & \int_{t-}^{t^+}\Big\|\Big[{N}(s),  \alpha^{S_n}_{s-t}(A(t))\Big]\Big\|\,ds\nonumber\\[2mm]
\label{ComIneq}
&=& \int_{t-}^{t^+}\Big\|\Big[\tilde{N}(s),  \alpha^{S_n}_{s-t}(A(t))\Big]\Big\|\,ds
\end{eqnarray}
where $t^-={\rm min}\{0,t\}$ and $t^+={\rm max}\{0,t\}$,
using the fact that  $\alpha^{S_n}_{s-t}(A(t))\in \al B._{S_n}$.
We will now estimate $\big\|\big[\tilde{N}(s),  \alpha^{S_n}_{s-t}(A(t))\big]\big\|$
(a method known as Lieb-Robinson bounds).

We first define the auxiliary function
\[
f(t):= \big[D,  \alpha^{S_n}_{t}(\beta_{-t}(A))\big]\in \al B._{S_m},
\]
 where $\beta_t:={\rm Ad}(e^{it(H_{S_n}^{\rm loc}+H_R^{\rm int})})$,
$A\in  \al B._R$, $R\subset S_n$, and $D$ is any bounded operator with  support
 in $S_m\backslash(S_n)_0$ (we have $\tilde{N}(s)$ in mind).
Then $\|f(0\|\leq 2\|A\|\|D\|\delta^{S_n}_R$
where  $\delta^{S_n}_R=0$ if  $R\cap S_m\backslash(S_n)_0=\emptyset$ (hence $\big[D,  A\big]=0$) and one otherwise.

We claim that $\beta_t$ preserves $\al B._R$. To see this, note that both $\alpha^{\rm loc}_{S_n}(t):={\rm Ad}(e^{itH_{S_n}^{\rm loc}})$
and $\alpha^{\rm int}_{R}(t):={\rm Ad}(e^{itH_R^{\rm int}})$ preserve $\al B._R$, hence restrict to
$\al K.(\al H._R)\subset \al B._R=\al B.(\al H._R)$. Then for these restrictions
$\alpha^{\rm loc}_{S_n}$ is a $C_0$-group (i.e. strongly  continuous), and $\alpha^{\rm int}_{R}$ is uniformly continuous
(with bounded generator). Then the sum of the generators of these two groups is again a generator of a $C_0$-group
on $\al K.(\al H._R)$ by   Theorem~3.1.33 in \cite{BR1}, and this $C_0$-group extends uniquely to an automorphic
action of $\R$ on $M(\al K.(\al H._R))$.  However the sum of the generators of the two groups on $\al K.(\al H._R)$ coincides
with the generator of the W*-action $\beta_t$, hence $\beta_t$ preserves $\al B._R$.


Now  $\alpha^{S_n}_{t}\circ\beta_{-t}={\rm Ad}\big(V(t,-t)\big)$ where
 $V(s,t):=\exp{(isH_{S_n})}\exp{(it(H_{S_n}^{\rm loc}+H_R^{\rm int}))}$. As  $V(s,t)$
 is strong operator continuous in $s,\,t$ and both
$H_{S_n}$ and $(H_{S_n}^{\rm loc}+H_R^{\rm int})$ have the same domain, which is
preserved by these unitaries, it follows from strong differentiation on the domain
and Lemma~\ref{Lemma1} in Appendix, that for all $\psi\in\cl H.$:
\begin{eqnarray*}
\frac{d}{dt}V(t,-t)\psi&=&ie^{itH_{S_n}}(H_{S_n}^{\rm int}-H_{R}^{\rm int})e^{-it(H_{S_n}^{\rm loc}+H_R^{\rm int})}\psi
=i(H_{S_n}^{\rm int}-H_{R}^{\rm int})(t)\,V(t,-t)\psi.
\\[2mm]
\hbox{Thus:}\qquad&&\\[2mm]
\frac{d}{dt}f(t)\psi&=&i\Big[D,  \alpha^{S_n}_{t}(\big[(H_{S_n}^{\rm int}-H_{R}^{\rm int}),\,\beta_{-t}(A)\big])\Big]\psi\\[2mm]
&=&i\Big[D,  \alpha^{S_n}_{t}\Big(\big[\sum_{q \in \Lambda^i_{S_n}
\backslash \Lambda^i_{R}}\Psi(q),
\,\beta_{-t}(A)\big]\Big)\Big]\psi\\[2mm]
&=&i\Big[D,  \alpha^{S_n}_{t}\Big(\big[\mathop{\mathord{\sum}'}_{q\in \Delta_{S_n}(R)} \Psi(q),
\,\beta_{-t}(A)\big]\Big)\Big]\psi
\end{eqnarray*}
where we used $\beta_{-t}(A)\in \al B._R$ in the third step.
Next, we define
\begin{eqnarray*}
\tilde{H}_{R}^{\rm int}&:=&\mathop{\mathord{\sum}'}_{q\in \Delta_{S_n}(R)} \Psi(q),\quad\hbox{then}\\[2mm]
\frac{d}{dt}V(t,-t)\psi &=&i\Big[D,  \alpha^{S_n}_{t}(\big[\tilde{H}_{R}^{\rm int},
\,\beta_{-t}(A)\big])\Big]\psi\\[2mm]
&=&i\big[\alpha^{S_n}_{t}(\tilde{H}_{R}^{\rm int}),f(t)\big]\psi
-i\Big[ \alpha^{S_n}_{t}(\beta_{-t}(A)),\big[ D,  \alpha^{S_n}_{t}(\tilde{H}_{R}^{\rm int})  \big]\Big]\psi.
\end{eqnarray*}
As the first term is a commutator with a bounded operator, it is a bounded linear map on $\al B._{S_m}$, hence by Lemma~\ref{Lemma3}
in the appendix we have
\begin{eqnarray*}
\|f(t\|&\leq&\|f(0)\|+2\|A\|\int_{t^-}^{t^+}\left\|  \big[ D,  \alpha^{S_n}_{r}(\tilde{H}_{R}^{\rm int})  \big] \right\|\,dr,
 \\[2mm]
\frac{\big\|\big[D,  \alpha^{S_n}_{t}(\beta_{-t}(A))\big]\big\|}{2\|A\|} &\leq &\|D\|\delta^{S_n}_R +
\mathop{\mathord{\sum}'}_{q\in\Delta_{S_n}(R)} \int_{t^-}^{t^+}\left\|  \big[ D,  \alpha^{S_n}_{r}(\Psi(q))  \big] \right\|\,dr.
\end{eqnarray*}
As $\beta_t:={\rm Ad}(e^{it(H_{S_n}^{\rm loc}+H_R^{\rm int})})$ preserves $\al B._R$, we can replace $A$
by $\beta_t(A)$ to get the estimate
\begin{equation}
 \label{CommToIterate}
\frac{\big\|\big[D,  \alpha^{S_n}_{t}(A)\big]\big\|}{2\|A\|} \leq \|D\|\delta^{S_n}_R+
\mathop{\mathord{\sum}'}_{q\in\Delta_{S_n}(R)} \int_{t^-}^{t^+}\left\|  \big[ D,  \alpha^{S_n}_{r}(\Psi(q))  \big] \right\|\,dr\,.
\end{equation}
If $\delta^{S_n}_R=0$, this inequality is potentially better than the naive inequality
\[
\big\|\big[D,  \alpha^{S_n}_{t}(A)\big]\big\|\leq 2\|A\|\|D\|
\]
which will be the case below when we let $S_n$ become large.
The inequality~$(\ref{CommToIterate})$ can now be iterated as $\Psi(q)\in \al B._{q}$.
If we substitute $\Psi(q)$ for $A$ in
$(\ref{CommToIterate})$ we get
\[
\frac{\big\|\big[D,  \alpha^{S_n}_{t}(\Psi(q))\big]\big\|}{2\|\Psi\|} \leq
\|D\|\delta^{S_n}_{q} +
\mathop{\mathord{\sum}'}\limits_{q'\in\Delta_{S_n}(q)}
 \int_{t^-}^{t^+}\left\|  \big[ D,  \alpha^{S_n}_{r}(\Psi(q'))  \big] \right\|\,dr.
\]
Substitution of this into the integrand of $(\ref{CommToIterate})$
and iterating produces:
\begin{eqnarray*}
\frac{\big\|\big[D,  \alpha^{S_n}_{t}(A)\big]\big\|}{2\|A\|} &\leq &\|D\|\delta^{S_n}_R+
\mathop{\mathord{\sum}'}_{q\in\Delta_{S_n}(R)}2\|D\|\|\Psi\|\delta^{S_n}_{q} \int_{t^-}^{t^+}1\,dr\\[1mm]
 +\;2\mathop{\mathord{\sum}'}_{q\in\Delta_{S_n}(R)}
   &&\!\!\!\!\!\!\!\!\!\!\!\!\!
 \mathop{\mathord{\sum}'}_{q'\in\Delta_{S_n}(q)}
 \|\Psi\|\int_{t^-}^{t^+}\int_{r^-}^{r^+}\left\|  \big[ D,  \alpha^{S_n}_{r'}(\Psi(q'))  \big] \right\|\,dr'\,dr\\[1mm]
 &\leq&\|D\|\Big(\delta^{S_n}_R+2\mathop{\mathord{\sum}'}_{q\in\Delta_{S_n}(R)}
 \|\Psi\|\delta^{S_n}_{q} |t|\\[1mm]
 &&\qquad\qquad+\;4\mathop{\mathord{\sum}'}_{q\in\Delta_{S_n}(R)}
 \mathop{\mathord{\sum}'}_{q'\in\Delta_{S_n}(q)}
 \|\Psi\|^2\delta^{S_n}_{q'} |t|^2/2\Big)\\[1mm]
 +\;4\mathop{\mathord{\sum}'}_{q\in\Delta_{S_n}(R)}\;\,
 \mathop{\mathord{\sum}'}_{q'\in\Delta_{S_n}(q)}
  &&\!\!\!\!\!\!\!\!\!\!\!\!
   \mathop{\mathord{\sum}'}_{q'' \in \Delta_{S_n}(q')}
 \|\Psi\|^2\int_{t^-}^{t^+}\int_{r_1^-}^{r_1^+}\int_{r_2^-}^{r_2^+}
 \left\|  \big[ D,  \alpha^{S_n}_{r_3}(\Psi(q''))  \big] \right\|\,dr_3\,dr_2\,dr_1\,.
\end{eqnarray*}
At the $N^{\rm th}$ iteration we have:
\begin{eqnarray*}
\frac{\big\|\big[D,  \alpha^{S_n}_{t}(A)\big]\big\|}{2\|A\|} &\leq &
\|D\|\Big(\delta^{S_n}_R+\mathop{\mathord{\sum}'}_{k=1}^N\frac{(2\|\Psi\||t|)^k}{k!}a_k\Big)
+R_N\\[1mm]
\hbox{where}\qquad a_k :=  \mathop{\mathord{\sum}'}_{q_1\in\Delta_{S_n}(R)}  &&\!\!\!\!\!\!\!\!\!\!\!\!\!
 \mathop{\mathord{\sum}'}_{q_2 \in\Delta_{S_n}(q_1)}\cdots
  \mathop{\mathord{\sum}'}_{q_k \in\Delta_{S_n}(q_{k-1})}
\delta^{S_n}_{q_k}                 \\[1mm]
R_N:=2^N\|\Psi\|^N\!\!\!\!\mathop{\mathord{\sum}'}_{q_1\in\Delta_{S_n}(R)}
\mathop{\mathord{\sum}'}_{q_2 \in\Delta_{S_n}(q_1)}
\!\!\!\!\!\!&\cdots&\!\!\!\!\!\!\!\!
\mathop{\mathord{\sum}'}_{q_{N+1}  \in\Delta_{S_n}(q_N) } \\[1mm]
\times &&
 \!\!\!\!\!\!\!\!\int_{t^-}^{t^+}\int_{r_1^-}^{r_1^+}\!\!\!\!\cdots\!\!\int_{r^-_{N-1}}^{r_{N-1}^+}
 \left\|  \big[ D,  \alpha^{S_n}_{r_{N+1}}(\Psi(q_{N+1}))  \big] \right\|\,dr_{N+1}\!\!\cdots dr_1\,.
\end{eqnarray*}
To prove that the iteration converges, we need to estimate the remainder term.  The integral is bounded by
$ 2\|\Psi\|\,\|D\| |t|^{N+1}/(N+1)!$
so we concentrate on counting the number of terms in the sums.

Let $S_n$ be the lattice cube with corner vertices $(\pm n,\pm n,\pm n)$ and $R=S_d$
for $d$ fixed, then for $d<n$ we have that as a link $\ell \in \Delta_{S_n}(R)$ must have one point in $R$
and one point outside, the pairs of endpoints of these links are
 uniquely specified by the points in $S_n\backslash S_d$ with
one nearest neighbour in $S_d$. For each of the six faces of $R=S_d$ there are $(2d+1)^2$ such points, hence
\[
|\Delta_{S_n}(R)\cap\Lambda^1|=6(2d+1)^2,
\]
where we ignored the fact that links have two possible orientations, because $\Lambda^1$ only contains one orientation for each link.
(We used the notation $|Z|$ for the cardinality of a set $Z$).

To estimate the number of plaquettes  $p\in \Delta_{S_n}(R)$, note that as at least one side of a  $p\in \Delta_{S_n}(R)$
must be a link $\ell \in \Delta_{S_n}(R)$, and given a link, there are 4 possible plaquettes it can belong to,
we get
\[
|\Delta_{S_n}(R)\cap\Lambda^2|\leq 4\times|\Delta_{S_n}(R)\cap\Lambda^1|= 24(2d+1)^2.
\]
We therefore have the estimate
\begin{equation}
\label{sumest}
\mathop{\mathord{\sum}'}_{q \in \Delta_{S_n}(R)}1
\leq 30(2d+1)^2.
\end{equation}
Next we want to estimate the size of $\Delta_{S_n}(q)\cap(\Lambda^1\cup\Lambda^2)$.
If $q$ is a link, then there are 10 links in $\Delta_{S_n}(q)\cap\Lambda^1$ for $n$ large enough,
and 20 plaquettes in $\Delta_{S_n}(q)\cap\Lambda^2$ (4 for which $q$ is a side, and 16 for which the
plaquette has only one vertex in common with $q$). If $q$ is a plaquette, then there are
16 links in $\Delta_{S_n}(q)\cap\Lambda^1$ for $n$ large enough, and
32 plaquettes in $\Delta_{S_n}(q)\cap\Lambda^2$ (8 in the plane of $q$, and 24 perpendicular to the plane of $q$).
Thus the number of terms in $\Delta_{S_n}(q)$ is less than or equal to the maximum of 30 and 48, i.e. 48.
Thus we have
\[
\|R_N\|\leq 30(2d+1)^2 (48)^N
\|D\|( 2\|\Psi\| |t|)^{N+1}/(N+1)!
\]
and it is clear for a fixed $t$ that this converges to $0$ as $N\to\infty$. We conclude that the iteration
converges. Thus
 \begin{equation}
\label{iterarg}
\frac{\big\|\big[D,  \alpha^{S_n}_{t}(A)\big]\big\|}{2\|A\|} \leq
\|D\|\Big(\delta^{S_n}_R+\sum_{k=1}^\infty\frac{(2\|\Psi\||t|)^k}{k!}a_k\Big).
 \end{equation}
Next, we want to estimate the coefficients $a_k$.
Assuming as above that $S_n$ is the lattice cube with corner vertices $(\pm n,\pm n,\pm n)$ and $R=S_d$
for $d$ fixed, then using the estimates above, we have
\[
a_1=\mathop{\mathord{\sum}'}_{q \in \Delta_{S_n}(R)} \delta^{S_n}_{q}
\leq 30(2d+1)^2.
\]
Recalling that
$\delta^{S_n}_R=0$ if  $R\cap S_m\backslash(S_n)_0=\emptyset$,
 if we keep $R$ fixed and let $n$ become large,
then $a_1=0$. 

We  can also use the estimates above for
\[
a_k =  \mathop{\mathord{\sum}'}_{q_1 \in \Delta_{S_n}(R)}
 \mathop{\mathord{\sum}'}_{q_2 \in \Delta_{S_n}(q_1)} \cdots
 \mathop{\mathord{\sum}'}_{q_k \in \Delta_{S_n}(q_{k-1}) }
 \delta^{S_n}_{q_k}
\]
Note that the sequence
\[
(q_1,q_2,\ldots,q_k)\quad\hbox{with}\quad q_i \in \Delta_{S_n}(q_{i-1})
\]
specifies a continuous path where the steps are either links or plaquettes,
starting from a $q_1\in \Delta_{S_n}(R)$ which has a point in $R$.
As $\Delta_{S_n}(S_r)\cap(\Lambda^1\cup\Lambda^2)\subset S_{r+1}$
and $\delta^{S_n}_{S_r}=0$ if $r<n-2$ we conclude that
$\delta^{S_n}_{q_k}=0$ whenever $k<n-d-2$.
Thus
\[
a_k =  \mathop{\mathord{\sum}'}_{q_1 \in \Delta_{S_n}(R)}
 \mathop{\mathord{\sum}'}_{q_2 \in \Delta_{S_n}(q_1)} \cdots
 \mathop{\mathord{\sum}'}_{q_k \in \Delta_{S_n}(q_{k-1}) }
 \delta^{S_n}_{q_k}
 \leq 30(2d+1)^2(48)^{k-1}\qquad\hbox{if}\quad k\geq n-d-2,
\]
and $a_k=0$ otherwise.
A substitution into  (\ref{iterarg})  produces:
\begin{eqnarray}
&&\!\!\!\!\!\!\!\!\!
\frac{\big\|\big[D,  \alpha^{S_n}_{t}(A)\big]\big\|}{2\|A\|} \leq
\|D\|\Big(\delta^{S_n}_{S_d}+\sum_{k=1}^\infty\frac{(2\|\Psi\||t|)^k}{k!}a_k\Big)\nonumber \\[1mm]
  &\leq&   \|D\|\Big(\delta^{S_n}_{S_d}+\sum_{k=n-d-2}^\infty\frac{(2\|\Psi\||t|)^k}{k!}
   30(2d+1)^2(48)^{k-1}\Big)               \nonumber      \\[1mm]
  \label{newLRs}
  &=&\|D\|(2d+1)^2\sum_{k=n-d-2}^\infty\frac{5(96\|\Psi\||t|)^k}{8(k!)}
  \qquad\hbox{if $n>d+4$.}
  \end{eqnarray}
 In order to estimate the integrand in (\ref{ComIneq}), we make the substitutions into Eq.~(\ref{newLRs})
 \[
 D\rightarrow \tilde{N}(s)={\rm Ad}\Big(e^{isH_{S_m\backslash S_n}^{\rm loc}}  \Big)\Big(\mathop{\mathord{\sum}'}_{q \in \Delta_{S_m}(
S_n)}\Psi(q)\Big),\qquad A\rightarrow A(t),\qquad t\rightarrow s-t,
 \]
where $s\in[t_-,t_+]$. Then we obtain
\[
\|\tilde{N}(s)\|=\Big\|\mathop{\mathord{\sum}'}_{q \in \Delta_{S_m}(
S_n)}\Psi(q)\Big\|\leq\|\Psi\|\mathop{\mathord{\sum}'}_{q \in \Delta_{S_m}(
S_n)}1\leq \|\Psi\|30(2n+1)^2
\]
by the estimates above. As it is obvious that $\|A(t)\|=\|A\|$ and $|s-t|\leq|t|$, we obtain  from
Eq.~(\ref{newLRs}) that if $n>d+4$ then
\begin{eqnarray*}
\big\|\big[\tilde{N}(s),  \alpha^{S_n}_{s-t}(A(t))\big]\big\| &\leq&
2\|\tilde{N}(s)\|\|A\|(2d+1)^2\sum_{k=n-d-2}^\infty
  \frac{5(96\|\Psi\||t|)^k}{8(k!)}\\[1mm]
&\leq& 2\|A\|\|\Psi\|30(2n+1)^2(2d+1)^2\sum_{k=n-d-2}^\infty
  \frac{5(96\|\Psi\||t|)^k}{8(k!)}\,.
\end{eqnarray*}
 Substitution of this into (\ref{ComIneq})gives for $n>d+4$:\\
\begin{eqnarray}
\label{NewCsumIneq}
&&\!\!\!\!\!\!\big\|\tau^{S_m}_t(A)-\tau^{S_n}_t(A)\|\nonumber\\[1mm]
&\leq&60\|A\|\|\Psi\|(2d+1)^2(2n+1)^2
\int_{t-}^{t^+}\sum_{k=n-d-2}^\infty\frac{5(96\|\Psi\||t|)^k}{8(k!)}\,ds\nonumber\\[1mm]
&=&\frac{75}{2}\|A\|\|\Psi\||t|(2d+1)^2(2n+1)^2\sum_{k=n-d-2}^\infty
  \frac{(96\|\Psi\||t|)^{k}}{k!}\nonumber\\[1mm]
&\leq&\frac{75}{192}\|A\|(2d+1)^2(2n+1)^2
  \frac{(96\|\Psi\||t|)^{n-d-1}}{(n-d-2)!}\exp\big(96\|\Psi\||t|\big)\,.
\end{eqnarray}
It is clear that this converges to zero as $n\to\infty$ for any  $t$.
Furthermore, by first taking the limit $m\to\infty$,
the estimate in (\ref{NewCsumIneq}) also shows that the limit in $n$ is uniform
for $t\in[-T,T]$ for a fixed $T$.
This concludes the proof of Theorem~\ref{GlobDynExist}.\\[3mm]

\begin{Lemma}
\label{alphadiscont}
With notation as above, we have for some $A\in\al A._{\rm max}$ that
$t\mapsto\alpha_t(A)$ is not norm continuous, i.e. $\alpha$ is not strongly continuous on
$\al A._{\rm max}$.
\end{Lemma}
\begin{beweis}
Assume the contrary, i.e. that $\alpha:\R\to{\rm Aut}(\al A._{\rm max})$ is strongly continuous. Then it has a densely defined
generator on $\al A._{\rm max}$, which we can perturb by a bounded generator $B$ to obtain a new
strongly continuous action $\alpha^B:\R\to{\rm Aut}(\al A._{\rm max})$. Fix $S\in\cl S.$ as a large enough lattice cube and recall
that $H_S= H_S^{\rm loc}+H_S^{\rm int}$ where $H_S^{\rm int}\in \al B._S\subset\al A._{\rm max}. $
Let the bounded perturbation $B$ then be
\[ B(A):=i[-H_S^{\rm int},A],\quad A \in \al A._{\rm max}. \]
Fix a strictly increasing sequence
$\{S_n\}_{n\in\N}\subset\cl S.$ such that ${S_n\nearrow\Z^3}$ as $n\to\infty$, and
$S\subseteq S_1$, and recall that $\alpha_t(A)=\lim\limits_{n\to \infty}\alpha^{S_n}_t(A)$. It  suffices to prove
that  $\alpha_t^B(A)=\lim\limits_{n\to \infty}\alpha^{S_n,B}_t(A)$ where
$\alpha^{S_n,B}_t:={\rm Ad}(\exp(it(H_{S_n}-H_S^{\rm int})))$. This is because on any subset $Z\subset S$ which cannot be reached
by a link or plaquette with a point on the boundary of $S$, we have that $\alpha^{S_n,B}_t$ (hence $\alpha_t^B$) coincides with
the free time evolution ${\rm Ad}(\exp(itH_Z^{\rm loc}))$.
As the free time evolution  preserves $\al B._Z\subset\al A._{\rm max}$ and is not strongly continuous on it,
this contradicts with the strong continuity of  $\alpha_t^B$.
Recall the Dyson series for bounded perturbations (cf. \cite[Theorem 3.1.33]{BR1}):
\begin{eqnarray*}
&&\!\!\!\!\!\!\!\!\alpha_t^B(A)=\alpha_t(A)+\\
&&\!\!\!\!\sum_{n=1}^\infty (-i)^n\int_0^t dt_1\int_0^{t_1}dt_2\cdots\int_0^{t_{n-1}}dt_n\big[
\alpha_{t_n}(H_S^{\rm int}),\big[\alpha_{t_{n-1}}(H_S^{\rm int}),\ldots[\alpha_{t_1}(H_S^{\rm int}),\alpha_t(A)]\ldots\big]\big].
\end{eqnarray*}
Now each multiple commutator in the integrands
\[
\big[
\alpha_{t_n}(H_S^{\rm int}),\big[\alpha_{t_{n-1}}(H_S^{\rm int}),\ldots[\alpha_{t_1}(H_S^{\rm int}),\alpha_t(A)]\ldots\big]\big]
\]
can be replaced by the norm limit
\[
\lim_{k\to \infty}\big[\alpha^{S_k}_{t_n}(H_S^{\rm int}),\ldots[\alpha^{S_k}_{t_1}(H_S^{\rm int}),\alpha^{S_k}_t(A)]\ldots\big]
\]
and as these norm limits are uniform on compact intervals in the time parameters, the limits can be taken through the
(weak operator convergent) integrals to produce:
\begin{eqnarray*}
\alpha_t^B(A)
&=&\lim_{k\to \infty}\Big\{\alpha^{S_k}_t(A)+\\
\sum_{n=1}^\infty\!\!\!\!\! &(-i)^n&\!\!\!\!\!\int_0^t dt_1\cdots\int_0^{t_{n-1}}dt_n
\big[\alpha^{S_k}_{t_n}(H_S^{\rm int}),\ldots[\alpha^{S_k}_{t_1}(H_S^{\rm int}),\alpha^{S_k}_t(A)]\ldots\big]\Big\}\\[2mm]
&=&\lim_{k\to \infty}\alpha^{S_k,B}_t(A)
\end{eqnarray*}
as required.
\end{beweis}

\subsection{Kinematics algebras and regular representations.}
\label{KARR}

In this subsection we will define our new kinematics algebra, show its relation to the
 kinematics field algebra we previously constructed in \cite{GrRu}, and consider
 the class of regular states and representations.

Having obtained the time evolution $\alpha:\R\to{\rm Aut}(\al A._{\rm max})$,
 we can now define our kinematics algebra as
\[
{\mathfrak A}_{\Lambda}:=C^*\Big(  \bigcup_{S\in\al S.}\alpha_{\R}({\mathfrak A}_S   )    \Big)
\subset\cl A._{\rm max}    \subset\cl B.(\cl H.).
\]
which is the minimal C*-algebra which contains all the local field algebras, and is preserved
by the dynamics.
Note that $\alpha$ does not preserve the local algebras.
For the full field algebra, one should take a crossed product w.r.t. the actions of desirable
transformations, such as gauge transformations (see below).
First, we want to show that the dynamics group $\alpha:\R\to{\rm Aut}({\mathfrak A}_{\Lambda})$
is in fact strongly continuous on ${\mathfrak A}_{\Lambda}$.
\begin{Theorem}
\label{GlobDynCont}
With notation as above, we have for all $A\in {\mathfrak A}_{\Lambda}$ that
$t\mapsto\alpha_t(A)$ is norm continuous, i.e. $\alpha$ is strongly continuous on
${\mathfrak A}_{\Lambda}$.
\end{Theorem}
\begin{beweis}
It suffices to prove continuity of
$t\mapsto\alpha_t(A)$
for $A\in{\mathfrak A}_R$ and $R\in\al S.$ arbitrary. Fix an $A\in{\mathfrak A}_R$, then
\[
\|\alpha_t(A)-A\|\leq\|\alpha_t(A)-\alpha^S_t(A)\|+\|\alpha^S_t(A)-A\|
\]
for any $S$. Fix a $\varepsilon>0$, then by Theorem~\ref{GlobDynExist} there is an
$S\in\al S.$ such that ${\|\alpha_t(A)-\alpha^S_t(A)\|}<\varepsilon/2$ for all
$t\in[-1,1]$, and it also holds for all larger $S$.
Fix such an $S$ such that $S\supset R$, 
and assume that $t\mapsto\alpha_t^S(A)$ is norm continuous (this will be proven below). So
$\lim\limits_{t\to 0}\alpha_t^S(A)=A$, hence there is a $\delta>0$ such that
$\|\alpha^S_t(A)-A\|<\varepsilon/2$ for all $t\in[-\delta,\delta]$. Thus if $|t|<\min\{1,\delta\}$,
we get that $\|\alpha_t(A)-A\|\leq\varepsilon$, i.e.
$\lim\limits_{t\to 0}\alpha_t(A)=A$ as required.

It remains to prove that $t\mapsto\alpha_t^S(A)$ is norm continuous for $A\in{\mathfrak A}_R$,
$R\subset S$
(note that $\alpha_t^S$ need not preserve ${\mathfrak A}_R$).
Let $\tau^{S}_t:={\rm Ad}(e^{itH_S}e^{-itH_S^{\rm loc}})$, then
\[
\alpha^{S}_t(A)=\tau^{S}_t\big( e^{itH_{S}^{\rm loc}} A e^{-itH_{S}^{\rm loc}} \big)
=\tau^{S}_t\big( e^{itH_R^{\rm loc}} A e^{-itH_R^{\rm loc}} \big) 
\]
hence it suffices to show that the maps $t\mapsto\tau^{S}_t(B)$ and 
$t\mapsto{\rm Ad}\big(e^{itH_R^{\rm loc}}\big)(A)=:A(t)$ are both norm continuous
for $A,\, B\in{\mathfrak A}_R$. For the map $t\mapsto A(t)$, recall that 
\[
e^{itH_R^{\rm loc}}=U^{\rm CAR}_R(t)\otimes
\bigotimes_{\ell\in\Lambda^1_R}U_\ell(t)\otimes\bigotimes\limits_{\ell'\not\in\Lambda^1_R}\1
\]
where  $U^{\rm CAR}_R(t)=\exp\big(itma^3 \sum\limits_{x \in \Lambda^0_R} \bar \psi_i (x)  \psi_i(x)\big)\in{\mathfrak F}_{R}$
and $U_\ell(t):=\exp(it\overline{E_{ij}(\ell) E_{ji}(\ell)})$. As the generator of $U^{\rm CAR}_R(t)$ is bounded, 
$t\mapsto U^{\rm CAR}_R(t)$ is norm continuous. As 
\[
{\mathfrak A}_R ={\mathfrak F}_R \otimes\mathop{\bigotimes}\limits_{\ell\in\Lambda^1_R}\al L._\ell
\otimes\mathop{\bigotimes}\limits_{\ell\not\in\Lambda^1_R}\un
\]
where $\al L._\ell:=C(G)\rtimes_\lambda G$
 acts on the factor  $\al H._\ell$ of $\al H._\infty$  as the algebra of compacts $\al K.(\al H._\ell)$
 and $t\mapsto U_\ell(t)KU_\ell(-t)$ is norm continuous for a compact operator $K\in\al K.(\al H._\ell)$,
 it follows that the map  $t\mapsto A(t)$ is norm continuous.

 Finally, as $t\mapsto\alpha^{S}_t\in{\rm Aut}(\al B._S)$ is a bounded perturbation of the weak operator continuous 
 one--parameter automorphism group $t\mapsto{\rm Ad}\big(e^{itH_S^{\rm loc}}\big)$ on $\al B._S\supset{\mathfrak A}_R$,
 we may express its cocycle  $\tau^{S}_t={\rm Ad}(e^{itH_S}e^{-itH_S^{\rm loc}})$ as a Dyson series:
 (cf. \cite[Theorem 3.1.33]{BR1}):
\begin{eqnarray*}
&&\!\!\!\!\!\!\!\!\tau^{S}_t(B)=B+\\[1mm]
&&\!\!\!\!\sum_{n=1}^\infty i^n\int_0^t dt_1\int_0^{t_1}dt_2\cdots\int_0^{t_{n-1}}dt_n\big[
H_S^{\rm int}(t_n),\big[H_S^{\rm int}(t_{n-1}),\ldots[H_S^{\rm int}(t_1),B]\ldots\big]\big],
\end{eqnarray*}
where $H_S^{\rm int}(t):={\rm Ad}\big(e^{itH_S^{\rm loc}}\big)(H_S^{\rm int})$ and thus
$\|H_S^{\rm int}(t)\|=\|H_S^{\rm int}\|<\infty$. Thus
we obtain the estimate
\[
\big\|\tau^{S}_t(B)-\tau^{S}_{t'}(B)\big\| 
\leq\sum_{n=1}^\infty \frac{\big|t^n-(t')^n \big|}{n!}
2^n\|H_S^{\rm int}\|^n\|B\|\,.
\]
It is clear that this convergent series approaches zero when $t\to t'$,
hence also $t\mapsto\tau^{S}_t(B)$ is continuous, from which it follows that 
 $t\mapsto\alpha_t^S(A)$ is norm continuous for $A\in{\mathfrak A}_R$, as required.
\end{beweis}
Thus  ${\mathfrak A}_{\Lambda}$  is  a convenient field algebra, in fact we can construct
for it the crossed product for the time evolution,
which is not possible for $\alpha:\R\to{\rm Aut}(\al A._{\rm max})$.

Next we want to examine covariant representations for
the automorphism group $\alpha:\R\to{\rm Aut}({\mathfrak A}_{\Lambda})$, and consider the question of
ground states. As the action $\alpha$ is a strongly continuous action of a locally compact group,
it defines a C*-dynamical system in the usual sense. Therefore we can construct its crossed product
(cf. \cite{Ped}),
and each representation of the crossed product produces a covariant representation for
$\alpha:\R\to{\rm Aut}({\mathfrak A}_{\Lambda})$. We therefore obtain a rich supply of covariant
representations. However, for physics, the physically appropriate class of representations should be regular in the following sense:
\begin{Definition}
A representation $\pi$ of ${\mathfrak A}_{\Lambda}$ is {\bf regular} if its restriction to
each local algebra ${\mathfrak A}_S$, $S\in \cl S.$, is nondegenerate (i.e. $\pi({\mathfrak A}_S)$
 has no nonzero
null spaces). A state is regular if its GNS representation is regular.
\end{Definition}
Note for a regular covariant representation $\pi$, that $\pi$ is also nondegenerate on
all the time evolved local algebras $\alpha_t({\mathfrak A}_S   )$.

The reasons why physical representations should be regular, are as follows. First, one requires that
the local observables do not all have zero expectation values w.r.t. any
(normalized) vector state in the representation,
i.e. the local observables in the field algebra should be visible in any
physically realizable state of the system.
Second, observe that the local Hamiltonians $H_S$ in the defining representation have compact resolvents
(see the next subsection below), hence  ${(i\un-H_S)^{-1}}\in {\mathfrak A}_S$. Thus, if a representation
 $\pi$ of ${\mathfrak A}_{\Lambda}$ is degenerate on ${\mathfrak A}_S$, then ${\pi((i\un-H_S)^{-1})}$ has a nonzero
 kernel, hence it cannot be the resolvent of an operator
 (cf.  Theorem 1 in \cite[p 216]{Yos}), thus the observable $H_S$ does not exist in this representation.
 Also observe, that this definition of regularity coincides with the one used in \cite{BuGr2, GrN09}.\\[3mm]

We now discuss the relation of the  kinematics field algebra we constructed before in \cite{GrRu}
with the kinematics field algebra ${\mathfrak A}_{\Lambda}$ we constructed here.
(For the full field algebras we need to add identities and construct
crossed products w.r.t. gauge transformations  and time evolutions).
The reader in a hurry may omit the rest of this subsection.

The kinematics field algebra constructed  in \cite{GrRu} is $\hat{\mathfrak A}_{\Lambda} :=
{\mathfrak F}_{\Lambda} \otimes {\cal L}[E]$ where  ${\mathfrak F}_{\Lambda}$ is the
Fermion algebra associated with $\Lambda$, and  $ {\cal L}[E]$ is a new ``infinite tensor product''
of the algebra of compact operators. Concretely,  $\hat{\mathfrak A}_{\Lambda}$
is represented on $\al H.=\al H._{\rm Fock}\otimes\al H._\infty$ as a tensor product representation as follows.
We let the Fermion algebra ${\mathfrak F}_{\Lambda}$ act in the Fock representation on $\al H._{\rm Fock}$.
Next, to see how $ {\cal L}[E]$ acts on $\al H._\infty$,
recall first that
 $\al H._\infty$  is the completion of the pre--Hilbert spanned by finite combinations of
elementary tensors of the type
\[
\varphi_1\otimes\cdots\otimes \varphi_k\otimes\psi_0\otimes\psi_0\otimes\cdots,\quad\varphi_i\in \al H._i= L^2(G),\;
k\in\N\,.
\]
Choose an increasing sequence of commuting finite dimensional projections $\{E_n\}_{n\in\N}\subset \cl K.\big(L^2(G)\big)
=\pi_0\big(C(G)\rtimes_\lambda G\big)$ which is an approximate identity for  $\cl K.\big(L^2(G)\big)$ and with
 $\psi_0\in E_nL^2(G)$ for all $n$. (By \cite{GrRu} there exists a choice of  $\{E_n\}_{n\in\N}$ which is
invariant w.r.t. a certain action of $G\times G$, but we will not insist on this point here.) Then
elementary tensors of the form
\[
A_1\otimes A_2\otimes\cdots\otimes A_k\otimes E_{n_{k+1}}\otimes E_{n_{k+2}}\otimes\cdots,\qquad
A_i\in\cl B.(\al H._i),\; n_j\in\N
\]
 act entrywise on  $\al H._\infty$ in a consistent way, so they define operators in
$\cl B.(\al H._\infty)$. In particular $ {\cal L}[E]$
is the C*-algebra generated by those elementary tensors where all $A_i\in\cl K.(\al H._i)$,
and we now have represented $\hat{\mathfrak A}_{\Lambda}$ on $\al H.$.
(In \cite{GrRu} we allowed the approximate identity  $\{E_n\}_{n\in\N}$ to be different for
different entries, but this generality is not needed).

For a finite connected subgraph $S\in\al S.$ of $(\Lambda^0, \Lambda^1)$, we  obviously
have the subalgebra $\hat{\mathfrak A}_S:={\mathfrak F}_{S}\otimes{\cal L}_{S}[E]$ where
${\cal L}_{S}[E]$ is the C*-algebra generated in $\cl B.(\al H._\infty)$ by those elementary tensors
of the type
\[
\Big(\mathop{\bigotimes}\limits_{\ell\in\Lambda^1_S}K_\ell\Big)\otimes
 \mathop{\bigotimes}\limits_{\ell\not\in\Lambda^1_S}E_\ell,\qquad K_\ell\in\cl K.(\cl H._\ell)
\]
where $E_\ell$ denotes an element of  $\{E_n\}_{n\in\N}$ placed in the position of the
tensor product corresponding to the link $\ell$.
As $\cl K.(\cl H._1)\otimes\cl K.(\cl H._2)=\cl K.(\cl H._1\otimes\cl H._2)$ we have in fact
that ${\cal L}_{S}[E]$ is the closure of the space spanned by
\[
\big\{K\otimes\mathop{\bigotimes}\limits_{\ell\not\in\Lambda^1_S}E_\ell\;\Big|\;K\in \cl K.(\mathop{\bigotimes}\limits_{\ell\in\Lambda^1_S}\cl H._\ell),\; E_\ell\in \{E_n\}_{n\in\N}\big\}.
 \]
 Thus by compactness of the projections $E_n$, we have $\hat{\mathfrak A}_S\subset \hat{\mathfrak A}_T$ if $S\subset T$.
 This should be contrasted with
  ${\mathfrak A}_S\subset M({\mathfrak A}_T)$ for the local algebras.
Now
 \[  \hat{\mathfrak A}_{\Lambda} =\ilim\hat{\mathfrak A}_S=\ilim\big({\mathfrak F}_{S}\otimes{\cal L}_{S}[E]\big).\]

Recall that for a  finite connected subgraph $S\in\al S.$ of $(\Lambda^0, \Lambda^1)$, we defined
\begin{eqnarray*}
\al H._S&:=& [{\mathfrak F}_{S}\Omega]\otimes[\al L._S\psi_0^\infty],\qquad\cl B._S:= \cl B.(\al H._S)\subset \cl B.(\al H.),\\[1mm]
\al A._{\rm max}&:=&\ilim \cl B._S = C^*\Big(\bigcup_{S\in\al S.}  \cl B.(\al H._S)   \Big),
\end{eqnarray*}
where an element of $\cl B.(\al H._S)$ acts as the identity on the factors $\al H._\ell$ of $\al H._\infty$ corresponding to
$\ell\not\in S$. By Equation~(\ref{BSisFB}), we realized $\cl B._S$ on $\cl H.$ by
\[
\cl B._S
=\pi_{\rm Fock}({\mathfrak F}_{S})\otimes\cl B.\Big(\mathop{\bigotimes}\limits_{\ell\in\Lambda^1_S}\al H._\ell\Big)
\otimes\mathop{\bigotimes}\limits_{\ell\not\in\Lambda^1_S}\un.
\]
The spanning elementary tensors  for $\hat{\mathfrak A}_{\Lambda}$ are all of the form
\[
A_F\otimes  K_S\otimes K_Q\otimes E_{n_{k+1}}\otimes E_{n_{k+2}}\otimes\cdots,\qquad A_F\in\pi_{\rm Fock}({\mathfrak F}_{R}),
\]
for $R=\{1,2,\ldots,k\}\supset S$, $ Q=R\backslash S $ and $K_\lambda\in
\cl K.\Big(\mathop{\bigotimes}\limits_{\ell\in\Lambda^1_\lambda}\al H._\ell\Big)$ for $\lambda\in\{S,Q\}$.
Thus the product of an elementary tensor in $\cl B._S$ with such a tensor can only change the first two factors,
and will again produce an elementary tensor of this kind. As the action of $\cl B.(\cl H.)$ on
$\cl K.(\cl H.)$ is nondegenerate, this implies that  $\cl B._S$ is in the multiplier algebra  of  $\hat{\mathfrak A}_{\Lambda}$.
Thus for the C*-algebra they generate we also have $\al A._{\rm max}\subset M\big(\hat{\mathfrak A}_{\Lambda}\big)$. As
 the kinematics field algebra ${\mathfrak A}_{\Lambda}$ we constructed here is contained in $\al A._{\rm max}$,
 we conclude that also ${\mathfrak A}_{\Lambda}\subset M\big(\hat{\mathfrak A}_{\Lambda}\big)$. This implies that every representation of
 $\hat{\mathfrak A}_{\Lambda}$ extends uniquely (on the same space) to  ${\mathfrak A}_{\Lambda}$, but not conversely.
 Every representation of ${\mathfrak A}_{\Lambda}$ which is obtained from a nondegenerate representation of  $\hat{\mathfrak A}_{\Lambda}$
 in this manner is regular, which is the content of the next lemma:-
\begin{Lemma}
\label{LemRegRep}
Given, in the notation above that ${\mathfrak A}_{\Lambda}\subset M\big(\hat{\mathfrak A}_{\Lambda}\big)$, let $\pi$ be
a nondegenerate representation of $\hat{\mathfrak A}_{\Lambda}$, and let $\tilde\pi$ be the unique extension of $\pi$
on the same Hilbert space to $M\big(\hat{\mathfrak A}_{\Lambda}\big)$. Then $\tilde\pi\restriction {\mathfrak A}_{\Lambda}$ is
regular. Likewise, if $\tilde\omega$ is the unique extension of a state $\omega$ on $\hat{\mathfrak A}_{\Lambda}$
 to a state on $M\big(\hat{\mathfrak A}_{\Lambda}\big)$, then
$\tilde\omega\restriction {\mathfrak A}_{\Lambda}$ is
regular.
\end{Lemma}
\begin{beweis}
By Theorem~A.2(vii) of \cite{GrN14}, it suffices for the first part to show that ${\mathfrak A}_S\cdot\hat{\mathfrak A}_{\Lambda}$
is strictly dense in $\hat{\mathfrak A}_{\Lambda}$ for all $S\in{\cal S}$. Recall from above that on $\cl H.$,
we have  ${\mathfrak A}_S={\mathfrak F}_{S}\otimes\al K.\Big(\mathop{\bigotimes}\limits_{\ell\in\Lambda^1_S}\al H._\ell\Big)
\otimes\mathop{\bigotimes}\limits_{\ell\not\in\Lambda^1_S}\un$, and that
$\hat{\mathfrak A}_{\Lambda}$ is spanned by the elementary tensors
\[
A_F\otimes  K_S\otimes K_Q\otimes E_{n_{k+1}}\otimes E_{n_{k+2}}\otimes\cdots,\qquad A_F\in\pi_{\rm Fock}({\mathfrak F}_{R}),
\]
for $R=\{1,2,\ldots,k\}\supset S$, $ Q=R\backslash S $ and $K_\lambda\in
\cl K.\Big(\mathop{\bigotimes}\limits_{\ell\in\Lambda^1_\lambda}\al H._\ell\Big)$ for $\lambda\in\{S,Q\}$.
Let $\{Y_i\}_{i\in\N}\subset\al K.\Big(\mathop{\bigotimes}\limits_{\ell\in\Lambda^1_S}\al H._\ell\Big)$ be an
approximate identity for $\al K.\Big(\mathop{\bigotimes}\limits_{\ell\in\Lambda^1_S}\al H._\ell\Big)$, then
$\un_F\otimes Y_i\otimes\mathop{\bigotimes}\limits_{\ell\not\in\Lambda^1_S}\un\in {\mathfrak A}_S$ where
$\un_F$ is the identity of ${\mathfrak F}_{R}$. Then
\begin{eqnarray*}
\Big(\un_F\otimes Y_i\otimes\mathop{\bigotimes}\limits_{\ell\not\in\Lambda^1_S}\un\Big)\Big(
A_F\otimes  K_S\otimes K_Q\otimes E_{n_{k+1}}\otimes E_{n_{k+2}}\otimes\cdots\Big)\\[1mm]
\qquad\qquad =A_F\otimes  Y_iK_S\otimes K_Q\otimes E_{n_{k+1}}\otimes E_{n_{k+2}}\otimes\cdots\qquad\qquad\qquad\\[1mm]
\qquad\qquad\qquad \longrightarrow A_F\otimes  K_S\otimes K_Q\otimes E_{n_{k+1}}\otimes E_{n_{k+2}}\otimes\cdots
\qquad\hbox{as}\; i\to\infty.
\end{eqnarray*}
Thus  ${\mathfrak A}_S\cdot\hat{\mathfrak A}_{\Lambda}$
is  dense in $\hat{\mathfrak A}_{\Lambda}$, and as the strict topology is weaker than the norm topology,
this implies strict density. As $S$ is arbitrary, this proves that
$\tilde\pi\restriction {\mathfrak A}_{\Lambda}$ is
regular.

If $\tilde\omega$ is the unique extension of a state $\omega$ on $\hat{\mathfrak A}_{\Lambda}$
 to a state on $M\big(\hat{\mathfrak A}_{\Lambda}\big)$, then $\pi_{\tilde\omega}$ is the
 unique extension of $\pi_\omega$
on the same Hilbert space to $M\big(\hat{\mathfrak A}_{\Lambda}\big)$. This is because $\tilde\omega$
is strictly continuous on $M\big(\hat{\mathfrak A}_{\Lambda}\big)$,
hence $\pi_{\tilde\omega}:M\big(\hat{\mathfrak A}_{\Lambda}\big)\to\cl B.(\cl H._{\tilde\omega})$
is strictly continuous w.r.t. the strong operator topology of $\cl B.(\cl H._{\tilde\omega})$.
By the strict density of
$\hat{\mathfrak A}_{\Lambda}$ in $M\big(\hat{\mathfrak A}_{\Lambda}\big)$ we get that
\[
{[\pi_{\tilde\omega}(M\big(\hat{\mathfrak A}_{\Lambda}\big))\Omega_{\tilde\omega}]}
={[\pi_{\tilde\omega}(\hat{\mathfrak A}_{\Lambda})\Omega_{\tilde\omega}]}
={[\pi_{\omega}(\hat{\mathfrak A}_{\Lambda})\Omega_{\omega}]}
=\cl H._\omega.
\]
Then $\pi_{\tilde\omega}$
is regular by the first part, hence $\tilde\omega\restriction {\mathfrak A}_{\Lambda}$
is regular.
\end{beweis}
By adapting the (lengthy) proof of Theorem 3.6 in \cite{GrN09}, we can also prove the converse, i.e.
that if a representation  of ${\mathfrak A}_{\Lambda}$ is regular, then it is obtained from a nondegenerate representation of 
$\hat{\mathfrak A}_{\Lambda}$ by the unique extension to $M\big(\hat{\mathfrak A}_{\Lambda}\big)$ 
on the same space. However, this will not be needed here.

\subsection{Ground states for the global dynamics.}
\label{PGDA}

Next we want to examine covariant representations for
the automorphism group ${\alpha:\R\to{\rm Aut}({\mathfrak A}_{\Lambda})}$.
 and consider the question of ground states.
For physics, only covariant representations where the generator of
time translations is positive is acceptable, and even more, for these representations ground states are needed.
As $\alpha:\R\to{\rm Aut}({\mathfrak A}_{\Lambda})$ is a continuous action of an amenable group,
it certainly has invariant states, but the difficult parts are to prove regularity and
the spectrum condition for
such an invariant state, establishing  it as a regular ground state.
To construct a ground state and establish
its properties, we will
follow a familiar method from \cite{BuGr2}.

First, we need to
consider the local Hamiltonians in greater detail.
As the restriction to $\cl H._S$ of the embedded copy of
$\cl B._S=\cl B.(\cl H._S)\subset\cl A._{\rm max}$ is faithful,
we will do the analysis on
\[
 \cl H._S=
  [{\mathfrak F}_{S}\Omega]\otimes[\al L._S\psi_0^\infty]=
[{\mathfrak F}_{S}\Omega]\otimes\mathop{\bigotimes}\limits_{\ell\in\Lambda^1_S}L^2(G)
 \otimes\bigotimes\limits_{\ell'\not\in\Lambda^1_S}\psi_0\subset\cl H..
\]
Separating the bounded and unbounded parts of $H_S$ on $\cl H._S$ we have:
\begin{eqnarray*}
H_S&=& H_S^{(0)}+H_S^{\rm bound}\qquad\hbox{on}\qquad
 [{\mathfrak F}_{S}\Omega]\otimes\tilde{\cl D.}_S\subset\cl H._S\subset
\cl H._{\rm Fock}\otimes\cl H._\infty\qquad\hbox{where:}\\[1mm]
 H_S^{(0)}  &:=& \tfrac{a}{2} \sum_{\ell \in \Lambda^1_S}
  E_{ij}(\ell) E_{ji}(\ell),\qquad H_S^{\rm bound}\in\cl B.(\cl H._S)\qquad\hbox{and}\\[1mm]
\tilde{\cl D.}_S&=&\bigotimes\limits_{\ell\in\Lambda^1_S}C^\infty(G)\otimes\bigotimes\limits_{\ell'\not\in\Lambda^1_S}\psi_0.
\end{eqnarray*}
Now $H_S^{(0)}=\un\otimes R_S\otimes\un$ where $R_S$ is the group Laplacian for the compact Lie group
 $G_S:=\prod\limits_{\ell\in\Lambda^1_S}G$ on
$L^2(G_S)\cong\bigotimes\limits_{\ell\in\Lambda^1_S}L^2(G)$. Thus by the theory of elliptic operators on compact
Riemannian manifolds, we conclude that for $R_S$ its eigenvalues are isolated, and its eigenspaces are finite dimensional,
cf. Theorem~III.5.8 in~\cite{LM89} and \cite{BH79}. Thus it has compact resolvent, i.e.
$(i\un-R_S)^{-1}\in\al K.(L^2(G_S))$. As $ [{\mathfrak F}_{S}\Omega]$ is finite dimensional, this is also true for
$H_S^{(0)}$ on $ [{\mathfrak F}_{S}\Omega]\otimes\tilde{\cl D.}_S\subset\cl H._S$, i.e.
$\big(i\un-H_S^{(0)}\big)^{-1}\in\al K.(\cl H._S)$. Then
\[
(i\un-H_S)^{-1}=\big(i\un-H_S^{(0)}\big)^{-1}+  (i\un-H_S)^{-1}  H_S^{\rm bound}\big(i\un-H_S^{(0)}\big)^{-1}\in\al K.(\cl H._S)
\]
hence $H_S=H_S^{(0)}+H_S^{\rm bound}$ also has discrete spectrum with finite dimensional eigenspaces. As $ H_S^{(0)}$ is positive and unbounded,
and $H_S^{\rm bound}$ is bounded, $H_S$ is bounded from below. Thus the lowest point in the spectrum of $H_S$ is an eigenvalue
$\lambda_S^{\rm grnd}\in\R$ with finite dimensional eigenspace $\cl E._S\subset\cl H._S$.
Fix a normalized
eigenvector  $\Omega_S\in\cl E._S\subset\cl H._S\subset\cl H.$. We conclude that the vector state
$\omega_S(\cdot):={\big(\Omega_S,\,\cdot\,\Omega_S\big)}$ is a ground state for the
local time evolution $\alpha^S:\R\to{\rm Aut}\big(\al A._{\rm max}\big)$ (and for its restriction to subalgebras such as
${\mathfrak A}_{\Lambda}$). In the original representation,
\[
\tilde{H}_S:=H_S-\un \lambda_S^{\rm grnd}
\]
will be the positive Hamiltonian for $\alpha^S_t={\rm Ad}\big(\exp(it\tilde{H}_S)\big)$ with smallest eigenvalue zero.

To construct a regular ground state for  $\alpha:\R\to{\rm Aut}({\mathfrak A}_{\Lambda})$ on ${\mathfrak A}_{\Lambda}$,
we proceed as follows.
Fix a strictly increasing sequence
$\{S_n\}_{n\in\N}\subset\cl S.$ such that ${S_n\nearrow\Z^3}$ as $n\to\infty$. For each $n\in\N$ choose
a state $\omega_n$ on $\cl B.(\cl H.)$
in the norm closed convex hull of vector states
\[
A\mapsto{\big(\Omega_{S_n},\, A\,\Omega_{S_n}\big)},\qquad A\in\cl B.(\cl H.),  
\]
where $\Omega_{S_n}\in\cl H._{S_n}$ ranges over the   normalized vectors in the eigenspace $\cl E._{S_n}$
 of $H_{S_n}$.
This sequence  $\{\omega_n\}_{n\in\N}$ need not converge, 
but by the Banach--Alaoglu theorem, the closed unit ball in $\cl B.(\cl H.)^*$
is compact in the  weak *--topology, hence the sequence $\{\omega_n\}_{n\in\N}$ has weak *--limit points,
and these limit points are states.
From such weak *--limit points we now want to show that we can obtain regular ground states
 on ${\mathfrak A}_{\Lambda}$. First we prove regularity.
\begin{Lemma}
\label{RegSt}
In the context above, for the  increasing sequence
 ${S_n\nearrow\Z^3}$ we fix  $S_n$ to be the lattice cube with corner vertices $(\pm n,\pm n,\pm n)$,
 which produces the sequence $\{\omega_n\}_{n\in\N}$. Let $\omega_\infty$ be a weak *--limit point
 of $\{\omega_n\}_{n\in\N}\subset\cl B.(\cl H.)^*$. Then
 the restriction of $\omega_\infty$ to ${\mathfrak A}_{\Lambda}\subset\cl B.(\cl H.)$ is regular.
\end{Lemma}
\begin{beweis}
We have to prove that $\pi^o_\infty({\mathfrak A}_S)$
is nondegenerate on $\al H.^o_\infty$ for all  $S\in\al S.$, where ${(\pi^o_\infty,\Omega^o_\infty,\al H.^o_\infty)}$
denotes the GNS--data of $\omega_\infty\restriction{\mathfrak A}_\Lambda$.

First consider $\omega_\infty$ on all of $\cl B.(\cl H.)$ with
 GNS--data ${(\tilde\pi_\infty,\tilde\Omega_\infty,\tilde{\al H.}_\infty)}$.
The GNS--data set of $\omega_\infty$ restricted to any subalgebra $\al A.\subset\cl B.(\cl H.)$ is just given by
the subspace ${[\tilde\pi_\infty(\al A.)\tilde\Omega_\infty]}\subset\tilde{\al H.}_\infty$ with the action of
$\tilde\pi_\infty(\al A.)$ on it, with cyclic vector $\tilde\Omega_\infty$. In particular, $\al H.^o_\infty$
is identified (i.e. unitarily equivalent) to
\[
[\tilde\pi_\infty({\mathfrak A}_\Lambda)\tilde\Omega_\infty]
=\Big[\tilde\pi_\infty\Big(C^*\Big(  \bigcup_{S\in\al S.}\alpha_{\R}({\mathfrak A}_S   )\Big)\Big)\tilde\Omega_\infty   \Big].
\]
We will prove below that  $\|\omega_\infty\restriction{\mathfrak A}_{S_n}\|=1$  for all $n$.
Assuming this, then on each $M({\mathfrak A}_{S_n})$,
 $\omega_\infty$ is  uniquely determined by its restriction to ${\mathfrak A}_{S_n}$
(cf. Prop.~2.11.7 in \cite{Dix}).
Let ${(\pi_\infty^{S_n},\Omega_\infty^{S_n},\al H._\infty^{S_n})}$
denote the GNS--data of $\omega_\infty\restriction M({\mathfrak A}_{S_n})$,
then this means that $\pi_\infty^{S_n}$ is strictly continuous w.r.t. the strong
operator topology of $\cl B.(\al H._\infty^{S_n})$.
Then, using the strict density of
${\mathfrak A}_{S_n}$ in $M({\mathfrak A}_{S_n})$, we obtain as in the proof of
Lemma~\ref{LemRegRep} that
\[
\left[\pi_\infty^{S_n}\big(M({\mathfrak A}_{S_n})\big)\Omega_\infty^{S_n}\right]
=\left[\pi_\infty^{S_n}\big({\mathfrak A}_{S_n}\big)\Omega_\infty^{S_n}\right]=\al H._\infty^{S_n}.
\]
In fact, using strict density of $C^*\Big(  \bigcup\limits_{S\subseteq S_n}\alpha_{\R}^{S_n}({\mathfrak A}_S   )    \Big)
\subseteq M({\mathfrak A}_{S_n})$, we have
\[
\Big[\pi_\infty^{S_n}\big(C^*\Big(  \bigcup_{S\subseteq S_n}\alpha_{\R}^{S_n}({\mathfrak A}_S   )
    \Big)\big)\Omega_\infty^{S_n}\Big]
=\left[\pi_\infty^{S_n}\big({\mathfrak A}_{S_n}\big)\Omega_\infty^{S_n}\right]=\al H._\infty^{S_n}.
\]
Using the identification above of $\al H._\infty^{S_n}$ with a subspace of
$\tilde{\al H.}_\infty$, this means that
\begin{eqnarray}
\label{pASnO}
\left[\tilde\pi_\infty\big(M({\mathfrak A}_{S_n})\big)\tilde\Omega_\infty\right]
&=&
\Big[\tilde\pi_\infty\big(C^*\Big(  \bigcup_{S\subseteq S_n}\alpha_{\R}^{S_n}({\mathfrak A}_S   )
    \Big)\big)\tilde\Omega_\infty\Big] \nonumber \\[1mm]
    \label{pASnO}
&=&\Big[\tilde\pi_\infty\big({\mathfrak A}_{S_n}\big)\tilde\Omega_\infty\Big]=\al H._\infty^{S_n}
\subset\tilde{\al H.}_\infty .
\end{eqnarray}
Note that  that if $n<m$ then $M({\mathfrak A}_{S_n})\subset M({\mathfrak A}_{S_m})$
hence equation~(\ref{pASnO}) implies that
\[
\Big[\tilde\pi_\infty\big({\mathfrak A}_{S_n}\big)\tilde\Omega_\infty\Big]\subseteq
\Big[\tilde\pi_\infty\big({\mathfrak A}_{S_m}\big)\tilde\Omega_\infty\Big].
\]
Moreover, by Theorem~\ref{GlobDynExist}
we have
$\alpha_t(A)=\lim\limits_{S\nearrow\Z^3}\alpha^S_t(A)$
hence $\alpha_t(A)$ for $A\in {\mathfrak A}_\Lambda$ is in the norm closure of
${\bigcup\limits_{n\in\N} \bigcup\limits_{S\subseteq S_n}\alpha_{\R}^{S_n}({\mathfrak A}_S   )}$ and so
from (\ref{pASnO}) we see
\begin{equation}
\label{piALO}
\al H.^o_\infty=
[\tilde\pi_\infty({\mathfrak A}_\Lambda)\tilde\Omega_\infty]
=\Big[\tilde\pi_\infty\Big(C^*\Big(  \bigcup_{S\in\al S.}\alpha_{\R}({\mathfrak A}_S   )\Big)\Big)\tilde\Omega_\infty   \Big]
 =\Big[\tilde\pi_\infty\Big(\bigcup_{n\in\N}{\mathfrak A}_{S_n} \Big)\tilde\Omega_\infty   \Big].
\end{equation}
Therefore, to prove  for a fixed $S$ that $\pi^o_\infty({\mathfrak A}_S)$
is nondegenerate on $\al H.^o_\infty$, it suffices to prove that it is nondegenerate on each of the spaces
${\big[\tilde\pi_\infty\big({\mathfrak A}_{S_n} \big)\tilde\Omega_\infty   \big]}$ as they are increasing in $n$,
and their union is dense in $\al H.^o_\infty$.
Let $k\in \N$ be large enough so that $S\subset S_k$ then for all $n\geq k$ we have that
${\mathfrak A}_S\subset M({\mathfrak A}_{S_n})$ and as ${\mathfrak A}_S$ acts nondegenerately on
${\mathfrak A}_{S_n}$, we have for any approximate identity $\{e_\gamma\}_{\gamma\in\Gamma}\subset{\mathfrak A}_S$
that $\lim\limits_\gamma e_\gamma A=A$ for all $A\in {\mathfrak A}_{S_n}$. Thus
\[
\lim_\gamma\big(\tilde\pi_\infty(e_\gamma)-\un\big)\big[\tilde\pi_\infty\big({\mathfrak A}_{S_n} \big)\tilde\Omega_\infty   \big]
=0\]
for all $n\geq k$, hence on all of $\al H._\infty$. Thus $\pi^o_\infty({\mathfrak A}_S)$
is nondegenerate on $\al H.^o_\infty$ for all $S\in\al S.$, i.e.
 $\omega_\infty$ restricted to ${\mathfrak A}_{\Lambda}\subset\cl B.(\cl H.)$ is regular.

It remains to prove that $\|\omega_\infty\restriction{\mathfrak A}_{S_n}\|=1$  for all $n$.
We will follow  the proof of  Lemma~7.3 in \cite{BuGr2}.
Let $m>n\in\N$ and on $\cl H._{S_m}\subset\cl H.$ consider
the operators $\tilde H_n:=\tilde H_{S_n},\;\tilde H_m:=\tilde H_{S_m}$ and $\tilde H_{m\backslash n}:=\tilde H_{S_m\backslash S_n}$,
and use analogous notation for $\lambda_m:=\lambda_{S_m}^{\rm grnd}$ etc.
As $[\tilde H_{m\backslash n},\tilde H_n]=0$ these operators have a joint dense domain on which $\tilde H_m$ is defined by
\[
\tilde H_m  = \tilde H_n + \tilde H_{m\backslash n} + \mathop{\mathord{\sum}'}_{q\in \Delta_{S_m}(S_n)} \Psi(q) +
(\lambda_n+\lambda_{m\backslash n}-\lambda_m)\un
\]
using notation from before in (\ref{primesum}) and (\ref{bdry}), as the additional terms are bounded. Let $\Omega$ be a normalized
joint eigenvector for $ \tilde H_n$ and $\tilde H_{m\backslash n}$ for the eigenvalue $0$, then
\[
0\leq(\Omega,\tilde H_m \Omega)=\Big(\Omega, \mathop{\mathord{\sum}'}_{q\in \Delta_{S_m}(S_n)} \Psi(q)\Omega\Big)
+\lambda_n+\lambda_{m\backslash n}-\lambda_m.
\]
Now recalling the estimate (\ref{sumest}), we have
\[
\mathop{\mathord{\sum}'}_{q \in \Delta_{S_m}(S_n)}\|\Psi\|
\leq 30(2n+1)^2\|\Psi\|.
\]
Thus the previous inequality gives
\[
\lambda_n+\lambda_{m\backslash n}-\lambda_m\geq -30(2n+1)^2\|\Psi\|,
\]
hence
\[
\tilde H_m + 60(2n+1)^2\|\Psi\|\un\geq \tilde H_n.
\]
Thus for all $\mu>0$ we have for the resolvents:
\[
\big(\tilde H_m + (\mu + 60(2n+1)^2\|\Psi\|)\un\big)^{-1}\leq (\tilde H_n +\mu\un)^{-1}\leq 1/\mu\,.
\]
From this we obtain
\[
 \mu\big(\mu + 60(2n+1)^2\|\Psi\|\big)^{-1}
 \leq \omega_m\big( \mu(\tilde H_n +\mu\un)^{-1}  \big)
 \leq 1.
\]
As $\omega_\infty$ is a weak *--limit point
 of $\{\omega_n\}_{n\in\N}\subset\cl B.(\cl H.)^*$, there is a subsequence  $\{\omega_{n_k}\}_{k\in\N}
 \subset\{\omega_n\}_{n\in\N}$  which converges to $\omega_\infty$ in the weak *--topology.
 We thus obtain
 \begin{equation}
\label{resolvineq}
 \mu\big(\mu + 60(2n+1)^2\|\Psi\|\big)^{-1}\leq\lim_{k\to\infty}\omega_{n_k}(\mu(\tilde H_n +\mu\un)^{-1})
 =\omega_\infty(\mu(\tilde H_n +\mu\un)^{-1})\leq 1.
  \end{equation}
Above we saw that on  $\cl H._{S_n}$ we have $(\tilde H_n +\mu\un)^{-1}\in\cl K.(\cl H._{S_n})$, hence
on $\cl H.$ we have
 \[
(\tilde H_n +\mu\un)^{-1}\in {\mathfrak A}_{S_n}={\mathfrak F}_{S_n}\otimes\al K.\Big(\mathop{\bigotimes}\limits_{\ell\in\Lambda^1_{S_n}}\al H._\ell\Big)
\otimes\mathop{\bigotimes}\limits_{\ell\not\in\Lambda^1_{S_n}}\un.
\]
This proves by (\ref{resolvineq}) that
$\|\omega_\infty\restriction{\mathfrak A}_{S_n}\|=1$  for all $n$.
\end{beweis}
\begin{Lemma}
\label{InvSt}
Assuming as above an  increasing sequence
 ${S_n\nearrow\Z^3}$ of lattice cubes, with  $\omega_\infty$  a weak *--limit point
 of $\{\omega_n\}_{n\in\N}\subset\cl B.(\cl H.)^*$, then $\omega_\infty\restriction\cl A._{\rm max}$
 is invariant w.r.t. the automorphism group $\alpha:\R\to{\rm Aut}(\cl A._{\rm max})$.
\end{Lemma}
\begin{beweis}
Let $\{\omega_{n_k}\}_{k\in\N}$ be a subsequence
weak *--converging to $\omega_\infty$. Observe that for any $A\in\cl A._{\rm max}$ and
$B\in\cl B.(\cl H.)$ that
\begin{eqnarray*}
\big|\omega_{n_k}(B\alpha^{S_{n_k}}_t(A)\! )&-&\!\omega_\infty(B\alpha_t(A))\big|\\[1mm]
&\leq& \big|\omega_{n_k}(B\alpha^{S_{n_k}}_t(A)-B\alpha_t(A))\big|
 +\big|\omega_{n_k}(B\alpha_t(A))-\omega_\infty(B\alpha_t(A))\big|\\[1mm]
&\leq & \|B\|\big\|\alpha^{S_{n_k}}_t(A)-\alpha_t(A)\big\|
+\big|\omega_{n_k}(B\alpha_t(A))-\omega_\infty(B\alpha_t(A))\big|\,.
\end{eqnarray*}
As the last expression goes to $0$ for $k\to\infty$, we get that
\begin{equation}
\label{lkioBS}
\lim\limits_{k\to\infty}\omega_{n_k}(B\alpha^{S_{n_k}}_t(A))=\omega_\infty(B\alpha_t(A)).
\end{equation}
Next observe that as the vector states $A\mapsto{\big(\Omega_{S_n},\, A\,\Omega_{S_n}\big)}$ are invariant
w.r.t. $\alpha^{S_n}$, so is any state in their norm closed convex hull, so $\omega_n$
is $\alpha^{S_n}\hbox{--invariant.}$
 Thus
\[
\omega_\infty(\alpha_t(A))=\lim_{k\to\infty}\omega_{n_k}(\alpha^{S_{n_k}}_t(A))
=\lim_{k\to\infty}\omega_{n_k}(A)=\omega_\infty(A)\,,
\]
hence $\omega_\infty$ is invariant for $\alpha$.
\end{beweis}

Consider the restriction of $\omega_\infty$  to ${\mathfrak A}_{\Lambda}\subset M\big(\hat{\mathfrak A}_{\Lambda}\big)$.
Denote the GNS--data of $\omega\restriction {\mathfrak A}_{\Lambda}$ by
${(\pi^o_\infty,\Omega^o_\infty,\al H.^o_\infty,U^o_\infty)}$ where
\[
U^o_\infty(t)\pi^o_\infty(A)\Omega^o_\infty=\pi^o_\infty(\alpha_t(A))\Omega^o_\infty\quad\forall\,A\in{\mathfrak A}_{\Lambda}.
\]
Then $(\pi^o_\infty, U^o_\infty)$ is a covariant pair for  $\alpha:\R\to{\rm Aut}( {\mathfrak A}_{\Lambda})$,
 in particular $t\mapsto U^o_\infty(t)$ is a weak operator continuous unitary group. This follows directly from the
 strong continuity of  $\alpha:\R\to{\rm Aut}({\mathfrak A}_{\Lambda})$
obtained in Theorem~\ref{GlobDynCont}.

Note that for 
  $\omega_\infty\restriction\cl A._{\rm max}$, the analogous 
statement need not be true,
because $\alpha:\R\to{\rm Aut}(\cl A._{\rm max})$ is not strongly continuous (cf. Lemma~\ref{alphadiscont}).
Finally, to establish that $\omega_\infty$ is a ground state, we need to prove that the generator of
$U^o_\infty(t)$ is nonnegative.
\begin{Theorem}
\label{groundState}
Assuming as above an  increasing sequence
 ${S_n\nearrow\Z^3}$ of lattice cubes, with  $\omega_\infty$  a weak *--limit point
 of $\{\omega_n\}_{n\in\N}\subset\cl B.(\cl H.)^*$, then $\omega_\infty\restriction{\mathfrak A}_{\Lambda}$
 is a regular ground state for  $\alpha:\R\to{\rm Aut}({\mathfrak A}_{\Lambda})$.
\end{Theorem}
\begin{beweis}
By the preceding lemmas, all that remains to be proven, is
that the generator of $U^o_\infty$ is nonnegative.
Let $h$ be a Schwartz function on $\R$, 
then we need to prove that  $U^o_\infty(h):=\int h(t)U^o_\infty(t)\,dt =0$ when ${\rm supp}(\hat{h})\subset (-\infty,0)$.
Fix a Schwartz function such that ${\rm supp}(\hat{h})\subset (-\infty,0)$.
By (\ref{piALO}) it suffices to prove  for all $A\in {\mathfrak A}_{S_n}$ and $n\in\N$ that
\[
0=  U^o_\infty(h) \pi^o_\infty(A)\Omega^o_\infty=\int h(t)U^o_\infty(t) \pi^o_\infty(A)\Omega^o_\infty\,dt
=  \int h(t) \pi^o_\infty(\alpha_t(A))\Omega^o_\infty\,dt.
  \]
 Let  $B\in{\mathfrak A}_{\Lambda}$ arbitrary, and let $\{\omega_{n_k}\}_{k\in\N}$ be a subsequence
weak *--converging to $\omega_\infty$. Then by equation (\ref{lkioBS}) we have
 \[
 \Big(\pi^o_\infty(B^*)\Omega^o_\infty,\, U^o_\infty(h) \pi^o_\infty(A)\Omega^o_\infty\Big) =
 \int h(t) \omega_\infty(B\alpha_t(A))\,dt
=\lim_{k\to\infty}\int h(t)\,\omega_{n_k}(B\alpha^{S_{n_k}}_t(A))\,dt
 \]
 using the dominated convergence theorem to take the limit through the integral.
 Observe that if $\omega_{n_k}$ is a vector state, i.e.
 $\omega_{n_k}(\cdot)={(\Omega_{S_{n_k}},\cdot\,\Omega_{S_{n_k}})}$ with $\Omega_{S_{n_k}}\in
 \cl E._{S_{n_k}}$,   then by
 $\tilde{H}_{S_{n_k}}\Omega_{S_{n_k}}=0$ and $\alpha^{S_{n_k}}_t={\rm Ad}\big(e^{it\tilde{H}_{S_{n_k}}}\big)$,
 we have:
\begin{eqnarray*}
\int h(t)\,\omega_{n_k}(B\alpha^{S_{n_k}}_t(A))\,dt
&=& \int h(t)\big(\Omega_{S_{n_k}},\,B e^{it\tilde{H}_{S_{n_k}}} A\,\Omega_{S_{n_k}}\big)\,dt\\[1mm]
&=&\big(\Omega_{S_{n_k}},\,B\int h(t) e^{it\tilde{H}_{S_{n_k}}}\,dt\,A\Omega_{S_{n_k}}\big)
=0
\end{eqnarray*}
where the strong operator integral $\int h(t) e^{it\tilde{H}_{S_{n_k}}}\,dt=0$
because $\tilde{H}_{S_{n_k}}\geq 0$ and ${\rm supp}(\hat{h})\subset (-\infty,0)$.
It follows that this also holds if $\omega_{n_k}$
is any state in the norm closed convex hull of these vector states.
Then
\[
\int h(t) \omega_\infty(B\alpha_t(A))\,dt
=\lim_{k\to\infty}\int h(t)\,\omega_{n_k}(B\alpha^{S_{n_k}}_t(A))\,dt=0\,.
\]
As $B$ is arbitrary, it follows that
$U^o_\infty(h) \pi^o_\infty(A)\Omega^o_\infty=0$ for all $A\in {\mathfrak A}_{S_n}$ and $n\in\N$
hence $U^o_\infty(h)=0$.
\end{beweis}
Note that there are several sources of nonuniqueness for ground states
in this argument. Apart from the possibility of different
 weak *--limit points of  $\{\omega_n\}_{n\in\N}$, there are also different choices of $\omega_n$ as
 the lowest eigenspace $\cl E._{S_n}$ of $H_{S_n}$ over which $\Omega_{S_n}$ ranges may have dimension higher than one.

\section{Gauge transformations and constraint enforcement.}
\label{GTGL}

To conclude this work, we need to define gauge transformations, enforce the Gauss law and
identify the physically observable subalgebra. This analysis was essentially done in \cite{GrRu}, but below we recall
the details for completeness. After enforcement of constraints, we will consider how the time evolution
automorphism group descends to the algebra of physical observables, and prove the existence of a
ground state for it.

By construction ${\mathfrak A}_{\Lambda} \subset\cl B.(\cl H.)$, hence to define gauge transformations
on ${\mathfrak A}_{\Lambda}$ it suffices to define a unitary representation of the gauge group on $\cl H.$
which implements the correct gauge transformations on the local subalgebras
${\mathfrak A}_S\subset{\mathfrak A}_{\Lambda}$.

The local gauge transformations
 on the lattice $\Lambda^0$ is
the group of maps $\gamma:\Lambda^0\to G$ of finite support, i.e.
\[
\gauc\Lambda := G^{(\Lambda^0)}=\big\{\gamma:\Lambda^0\to G\,\mid\,\big|{\rm supp}(\gamma)\big|<\infty\big\},\qquad
{\rm supp}(\gamma):=\{x\in\Lambda^0\,\mid\,\gamma(x)\not=e\}.
\]
This is an inductive limit indexed by the finite subsets $S\subset\cl S.$,
of the subgroups ${\rm Gau}_S \Lambda:=\{\gamma:\Lambda^0\to G\,\mid\,{\rm supp}(\gamma)\subseteq S\}
\cong\prod\limits_{x\in S}G$, and we give it the inductive limit
topology.
It acts on each local field algebra
 ${\mathfrak A}_S={\mathfrak F}_{S}\otimes\mathop{\bigotimes}\limits_{\ell\in\Lambda^1_S}\cl L._\ell$
 by a product action as above in Sect.~\ref{PM}, implemented by a unitary (cf.~(\ref{LocalW}))
\[
\hat{W}_\zeta:=U^F_\zeta\otimes\big(\bigotimes_{\ell\in\Lambda^1_S}W^{(\ell)}_\zeta\big),\quad
\zeta\in\gauc \Lambda_S,
\]
on  $\cl H._F\otimes\mathop{\bigotimes}\limits_{\ell\in\Lambda^1_S}L^2(G)$. Here $U^F_\zeta$ is the second quantization
on the fermionic Fock space  $\cl H._F$ of the transformation $f\to \zeta\cdot f$ where
$(\zeta\cdot f)(x):={\zeta(x)}f(x)$ for all $x\in\Lambda_S^0$ and $f\in\Cn_S$, hence $U^F_\zeta$ implements the automorphism
$\alpha^1:\gauc \Lambda_S \to\aut{\mathfrak F}_{\Lambda_S}$ by
$\alpha_\zeta^1(a(f)):=a(\zeta\cdot f)$.
The  $W^{(\ell)}_\zeta$ are copies of the unitaries $W_\zeta:L^2(G)\to L^2(G)$ by
\[
(W_\zeta\varphi)(h):=\varphi(\zeta^{-1}\cdot h)=\varphi(\zeta(x_\ell)^{-1}\,h\,\zeta(y_\ell)).
\]
For the full infinite lattice, these unitaries generalize naturally to
$\al H.:=\al H._{\rm Fock}\otimes\al H._\infty$ by the same formulae, as each
$\zeta\in\gauc \Lambda$ is of finite support, i.e.
\[
\hat{W}_\zeta:=U^F_\zeta\otimes\big(\bigotimes_{\ell\in\Lambda^1_{{\rm supp}'(\gamma)}}W^{(\ell)}_\zeta\big),\quad
\zeta\in\gauc \Lambda
\]
where $U^F_\zeta$ is again the second quantization
on the fermionic Fock space  $\al H._{\rm Fock}$ of the map $f\to\zeta\cdot f$.
Here ${\rm supp}'(\gamma)$ denotes the
subgraph of $\Lambda$ consisting of all the links which have at least one point in
${\rm supp}(\gamma)$.
Hence
$\Lambda^1_{{\rm supp}'(\gamma)}$ consists of the links which have at least one point in
${\rm supp}(\gamma)$.
In this notation, we assumed that $\hat{W}_\zeta$ acts as the identity
on those factors of $\cl H._\infty$ corresponding to $\ell\not\in{\rm supp}'(\gamma)$.

This produces a unitary representation $\hat{W}:\gauc \Lambda\to\al U.(\al H.)$.
Then the gauge transformation
produced by $\zeta$ on the system of operators is given by ${\rm Ad}(\hat{W}_\zeta)$, and
on the local algebras it produces the same gauge transformations as in Subsect.~\ref{PM}.
It is clear also that these gauge transformations preserve
 \[
 \al A._{\rm max}=\ilim \cl B._S = C^*\Big(\bigcup_{S\in\al S.}  \cl B.(\al H._S)   \Big)
\]
hence we can use  ${\rm Ad}(\hat{W}_\zeta)$ to define gauge transformations on our maximal algebra.

Next we recall that the local Hamiltonians $H_S$ are constructed from gauge invariant terms,
and hence  ${\rm Ad}(\hat{W}_\zeta)(e^{itH_S})=e^{itH_S}$ and so for
$\alpha_t^S:={\rm Ad}(e^{itH_S})$ we have ${\alpha_t^S\circ {\rm Ad}(\hat{W}_\zeta)}={\rm Ad}(\hat{W}_\zeta)\circ\alpha_t^S$.
Thus for the global time evolutions on $ \al A._{\rm max}$ we get from Theorem~\ref{GlobDynExist}
that for all $A\in \al A._{\rm max}$ and $\zeta\in\gauc \Lambda$ we have
\begin{eqnarray*}
\hat{W}_\zeta\,\alpha_t(A)\,\hat{W}_\zeta^*&=& \hat{W}_\zeta\Big(\lim_{S\nearrow\Z^3}\alpha^S_t(A)\Big)\hat{W}_\zeta^*
=\lim_{S\nearrow\Z^3} \hat{W}_\zeta\alpha^S_t(A) \hat{W}_\zeta^*\\[1mm]
&=&\lim_{S\nearrow\Z^3}\alpha^S_t\big(\hat{W}_\zeta A\hat{W}_\zeta^* \big)
=\alpha_t\big(\hat{W}_\zeta A\hat{W}_\zeta^* \big)
\end{eqnarray*}
i.e. the global time evolution also commutes with the gauge transformations. This implies that the gauge transformations
${\rm Ad}(\hat{W}_\zeta)$ will preserve all orbits of the global time evolution, and hence ${\rm Ad}(\hat{W}_\zeta)$
preserves our kinematics algebra ${\mathfrak A}_{\Lambda}:=C^*\Big(  \bigcup_{S\in\al S.}\alpha_{\R}({\mathfrak A}_S   )    \Big)$.
By restriction, the gauge transformations are therefore well-defined  on ${\mathfrak A}_{\Lambda}$,
and we will denote the action by $\beta:\gauc \Lambda\to{\rm Aut}({\mathfrak A}_{\Lambda})$.
Note that using ${\mathfrak A}_{\Lambda}\subset M(\hat{\mathfrak A}_{\Lambda})$ this action is the
canonical extension of the one on $\hat{\mathfrak A}_{\Lambda}$ defined in  \cite{GrRu}.

Finally, we would like to enforce the Gauss law constraint and
identify the physical subalgebra. In \cite{GrRu} we already did this for the local algebras ${\mathfrak A}_S$ in
Theorem~4.12, and proved in Theorem~4.13 that the traditional constraint enforcement method - taking the gauge invariant part of the algebra,
then  factoring out the residual constraints -
produced results coinciding with those of the T--procedure (cf.~\cite{GrSrv}).
As the time evolution automorphism group commutes with the gauge transformations, it will
respect the constraint reduction, hence define a time evolution automorphism group on the
algebra of physical observables.

The concrete constraint reduction in the defining representation of  ${\mathfrak A}_{\Lambda} \subset\cl B.(\cl H.)$
starts with the representation $\hat{W}:\gauc \Lambda\to\al U.(\al H.)$ of the gauge group which implements the
gauge transformations. One defines the gauge invariant subspace
\[
\al H._G:=\{\psi\in\al H.\,\mid\,\hat{W}_\zeta\psi=\psi\;\;\forall\,\zeta\in\gauc \Lambda\}
\]
and observes that the cyclic vector $\Omega\otimes\psi_0^\infty$ is in $\al H._G$. Let $P_G$ be the projection
onto $\al H._G$, then an $A\in\cl B.(\cl H.)$ commutes with  $P_G$ iff both $A$ and $A^*$ preserve $\al H._G$.
Thus our observables are in
\[
\{P_G\}'\cap {\mathfrak A}_{\Lambda} \subseteq P_G \cl B.(\cl H.)P_G +  (\un-P_G) \cl B.(\cl H.)(\un-P_G).
\]
The final step of constraining consists of restricting $\{P_G\}'\cap {\mathfrak A}_{\Lambda}$ to
$\al H._G$, which means that we discard the second part of the decomposition above. On the local algebras
$ {\mathfrak A}_S$ this will produce a copy of the algebra of compact operators  on the gauge invariant part of $\al H._S$
by Theorem~4.13 in \cite{GrRu}. The algebra generated by the orbit of the time evolutions of these reduced local algebras
will be a particularly important subalgebra of algebra of physical observables, as it is constructed purely from
the original physical observables with no involvement of the gauge variables.\\[2mm]

\noindent{\bf Remark:}\\
The terminology we use here comes from  the T--procedure (cf.~\cite{GrSrv}), where the algebra
consistent with the constraints is called the observable algebra (here it is
 $\{P_G\}'\cap {\mathfrak A}_{\Lambda}$),  and the final algebra obtained from it by factoring out the ideal generated by
 the constraints is called the algebra of physical observables. This is different from the terminology in \cite{KR, JKR}
 where the observable algebra is the final algebra obtained by factoring out the ideal generated by
 the constraints from the algebra of gauge invariant variables.\\[2mm]

To conclude this section, we will next prove that there are gauge invariant ground states.
By Proposition~4.2 in \cite{GrRu} these produce Dirac states on the original algebra,
hence states on the algebra of physical observables (cf. Theorem~4.5 in \cite{GrRu}).
Such a  state on the algebra of physical observables  is a ground state w.r.t. the
time evolution automorphism group on the algebra of physical observables which is
descended from the one on ${\mathfrak A}_{\Lambda}$.
\begin{Theorem}
There is a gauge invariant regular ground state for  $\alpha:\R\to{\rm Aut}({\mathfrak A}_{\Lambda})$.
\end{Theorem}
\begin{beweis}
Recall from Theorem~\ref{groundState}, that if we take
an  increasing sequence
 ${S_n\nearrow\Z^3}$ of lattice cubes, with  $\omega_\infty$  a weak *--limit point
 of $\{\omega_n\}_{n\in\N}\subset\cl B.(\cl H.)^*$, then $\omega_\infty\restriction{\mathfrak A}_{\Lambda}$
 is a regular ground state for  $\alpha:\R\to{\rm Aut}({\mathfrak A}_{\Lambda})$.
Here each $\omega_n$ is in the closed convex hull
of vector states
\[
A\mapsto{\big(\Omega_{S_n},\, A\,\Omega_{S_n}\big)},\qquad A\in\cl B.(\cl H.)
\]
and $\Omega_S\in\cl H._S$ ranges over the   normalized
eigenvectors  of the lowest point in the spectrum of $H_S$. It is clear that if the sequence
 $\{\omega_n\}_{n\in\N}\subset\cl B.(\cl H.)^*$ consists of gauge invariant
 states, then so are its weak *--limit points, hence it suffices to show for any $S_n\in\al S.$ that we can find
 $\omega_n$ chosen as above, which is invariant w.r.t. conjugation by
 the unitaries $\hat{W}:\gauc \Lambda\to\al U.(\al H.)$.

 As $\hat{W}(\gauc \Lambda)$ commutes with each  $H_S$, it leaves its eigenspaces invariant,
and
 $\hat{W}(\gauc \Lambda)$ restricted to $\al H._S$ is just  $\hat{W}({\rm Gau}_S \Lambda )$.
  This group is  compact, hence by the Peter-Weyl theorem, each eigenspace of  $H_S$
  in  $\al H._S$ decomposes into finite dimensional subspaces on which
  $\hat{W}({\rm Gau}_S \Lambda )$ acts irreducibly. Choose for the lowest eigenspace 
   $\cl E._S\subset\al H._S$ of 
  $H_S$   such an irreducible component of $\hat{W}({\rm Gau}_S \Lambda )$
  contained in it, and denote the finite dimensional component space by $V_S\subseteq\cl E._S$.
  Denote the unit sphere of  $V_S$ by $E_S$. By finite dimensionality of  $V_S$,
  $E_S$ is compact. Consider the map $\eta:\al H.\to\al B.(\al H.)^*$ by
  $\eta(\psi)(A):={(\psi,A\psi)}$ for $\psi\in\al H.$, $A\in\al B.(\al H.)$.
  Then $\eta$ is continuous w.r.t. the Hilbert space topology and the w*-topology of
  $\al B.(\al H.)^*$, hence $\eta$ takes compact sets to compact sets. In particular
  $\eta(E_S)\subset{\got S}(\al B.(\al H.))$ (denoting the state space of $\al B.(\al H.)$),
  is compact in the w*-topology. Denote the
  norm closed convex hull of  $\eta(E_S)$ by ${\got S}_S$, and observe that it is
  contained in the finite dimensional subspace of $\al B.(\al H.)^*$ spanned by the
  functionals $A\mapsto{(v_i,Av_j)}$, $A\in\al B.(\al H.)$, where $v_i$ and $v_j$ range over
  some orthonormal basis $\{v_1,\ldots,v_k\}$ of $V_S$. As all Hausdorff vector topologies of a
  finite dimensional vector space coincide (cf. \cite[Theorem~I.3.2]{Sch}), we conclude
  from norm boundedness and closure, that
  ${\got S}_S$ is compact w.r.t. the w*-topology.


  As $E_S$ is invariant as a set w.r.t. the action of
  $\hat{W}({\rm Gau}_S \Lambda )$, we conclude that $\eta(E_S)$ is invariant w.r.t.
  the action of ${\rm Gau}_S \Lambda$ on $\al B.(\al H.)^*$ by
  \[
  \varphi\mapsto\varphi_\zeta,\quad\hbox{where}\quad\varphi_\zeta(A):=\varphi(
  \hat{W}_\zeta A \hat{W}_\zeta^*)
  \]
for $\zeta\in {\rm Gau}_S \Lambda$, $A\in \al B.(\al H.)$ and $\varphi\in \al B.(\al H.)^*$.
As the action comes from automorphisms on $\al B.(\al H.)$, it extends to an affine
action on  the closed convex hull  ${\got S}_S$, which is w*-compact.
This action is continuous in the w*-topology, hence as ${\rm Gau}_S \Lambda$ is
amenable (as it is compact), we obtain that there is an invariant point
$\hat\omega_S\in{\got S}_S$ (cf.  \cite[Theorem~5.4]{Pie}).
However, the action of ${\rm Gau}_S \Lambda$ on ${\got S}_S$ is just the restriction of the
action of ${\rm Gau} \Lambda$ on ${\got S}(\al B.(\al H.))$ to ${\got S}_S$,
hence $\hat\omega_S\in{\got S}_S$ is gauge invariant, and can be restricted to
${\mathfrak A}_{\Lambda}$.
This concludes the proof that there are gauge invariant ground states.
\end{beweis}

\section{Conclusions.}

In the preceding, we have proven the existence of the dynamical automorphism group for
QCD on an infinite lattice, and obtained a suitable minimal field algebra on which it acts.
This pair defines a C*-dynamical system in the sense that it is strongly continuous.
We proved the existence of regular ground states, and discussed how to enforce the Gauss law constraint.

Clearly much more remains to be done, e.g
\begin{itemize}
\item[(1)] There is no uniqueness proven or analyzed for the ground states. One needs to determine
the properties of the set of ground states.
\item[(2)] The form and structure of the physical observable algebra needs to be determined more explicitly.
\item[(3)] Existence of the dynamics is not enough, some useful approximation schemes are needed to connect
with present analysis on finite lattices.
\end{itemize}

\section{Appendix}

\begin{Lemma}
\label{Lemma1}
If $B,\; V:\R\to\al B.(\cl H.)$ are bounded strong operator continuous maps, and satisfy
\begin{equation}
\label{ddtVA}
\frac{d}{dt}V(t)\psi=B(t)\psi\qquad\forall\,t\in\R
\end{equation}
for all $\psi$ in some dense subspace $\al D.$ of $\al H.$, then the relation
(\ref{ddtVA}) holds for all
$\psi\in\al H.$.
\end{Lemma}
\begin{beweis}
A version of this lemma is proven in Prop.~2.1 (iii) in~\cite{NaSi2}, but as that proof is indirect, we give a direct proof here.
We will use the notation $\|V\|:=\sup\limits_{t\in\R}\|V(t)\|$ and $\|B\|:=\sup\limits_{t\in\R}\|B(t)\|$. By integration of
(\ref{ddtVA}) we obtain for all $\psi\in\al D.$ and $h\not=0$ that
\begin{eqnarray*}
\big(V(t+h)-V(t)\big)\psi &=& \int_t^{t+h}B(s)\psi\,ds  \qquad\forall\,\psi\in\al D.,\qquad\hbox{and hence:}\\[1mm]
\Big\|\frac{1}{h}\big(V(t+h)-V(t)\big)\psi\Big\| &\leq& \frac{1}{|h|}\int_{t_-}^{t_+}\|B(s)\psi\|\,ds
\leq\|B\|\,\|\psi\|
\end{eqnarray*}
where $t_-:=\min(t, t+h)$ and $t_+:=\max(t, t+h)$. As $\frac{1}{h}\big(V(t+h)-V(t)\big)$ for fixed $t,\,h$ is bounded,
and $\al D.$ is dense, this implies that $\|\frac{1}{h}\big(V(t+h)-V(t)\big)\|\leq\|B\|.$ Let
\[
D_h:=\frac{1}{h}\big(V(t+h)-V(t)\big)-B(t)\quad\hbox{hence}\quad \|D_h\|\leq 2\|B\|.
\]
We want to prove that $\lim\limits_{h\to 0} D_h\varphi=0$ for all $\varphi\in\al H.$, i.e. that
(\ref{ddtVA}) holds on all of $\al H.$. Let $\varphi\in\al H.$ and choose a sequence $\psi_n\in\al D.$
such that $\psi_n\to\varphi$. Fix an $\varepsilon >0$. For any $n\in\N$:
\[
\|D_h\varphi\|\leq\|D_h(\varphi-\psi_n)\|+\|D_h\psi_n\|\leq 2\|B\|\,\|\varphi-\psi_n\|+\|D_h\psi_n\|.
\]
Choose an $n$ such that $2\|B\|\,\|\varphi-\psi_n\|<\varepsilon/2$. By the limit in (\ref{ddtVA}) there is a $\delta>0$
such that $|h|<\delta$ implies $\|D_h\psi_n\|<\varepsilon/2$, and hence $\|D_h\varphi\|<\varepsilon$. Thus
$\lim\limits_{h\to 0} D_h\varphi=0$ as required.
\end{beweis}
\begin{Lemma}
\label{Lemma2}
If $A:\R\to\al B.(\cl H.)$ is a measurable map w.r.t. the strong operator topology, and $\cl H.$ is separable, then
$A:\R\to\al B.(\cl H.)$ is a measurable map w.r.t. the norm topology. Then for any bounded interval $I$ we have
\[
\Big\|\int_I A(t)\, dt\Big\|\leq \int_I\| A(t)\| dt.
\]
\end{Lemma}
\begin{beweis}
(cf. Eq.~(12) in~\cite{NaSi2})\\
By assumption, $t\mapsto\|A(t)\psi\|$ is measurable for each $\psi\in\cl H.$. Consider the supremum
\[
\|A(t)\|=\sup\{\|A(t)\psi\|\,\mid\,\psi\in\cl H.,\;\|\psi\|\leq 1\}.
\]
As $\cl H.$ is separable, there is a countable dense set in the closed unit ball of $\cl H.$, which we can arrange
into a sequence $(\psi_n)_{n\in\N}$ and hence obtain
\[
\|A(t)\|=\sup\{\|A(t)\psi_n\|\,\mid\,n\in\N\}.
\]
However, the supremum of a sequence of measurable functions is measurable, hence $t\mapsto\|A(t)\|$ is measurable.
For the Bochner integral of $t\mapsto A(t)\psi\in\cl H.$ we thus obtain
\[
\Big\|\Big(\int_I A(t)\, dt\Big)\psi\Big\|=\Big\|\int_I A(t)\psi\, dt\Big\|\leq
\int_I\| A(t)\psi\| dt\leq \int_I\| A(t)\| dt\cdot\|\psi\|.
\]
By taking the supremum over $\psi$ in the closed unit ball of $\cl H.$ we obtain that\\[1mm]
$\Big\|\int_I A(t)\, dt\Big\|\leq \int_I\| A(t)\| dt.$
\end{beweis}

\begin{Lemma}
\label{Lemma3}
Let $A,\; B:\R\to\al B.(\cl H.)$ be  strong operator continuous maps, such that
$A(t)^*=A(t)$ and $\|A(t)\|<M$ for all $t$ for a fixed $M$, and assume that $\cl H.$
is separable. Then for any $t_0\in\R$ and
$f_0\in\al B.(\cl H.)$,  there is a unique
strong operator differentiable map $f:\R\to\al B.(\cl H.)$ such that
\[
\frac{d}{dt}f(t)\psi = i[f(t),A(t)]\psi+B(t)\psi\quad\forall\,\psi\in\cl H.,\quad
\hbox{and}\quad f(t_0)=f_0\in\al B.(\cl H.).
\]
This solution $f$ of the IVP satisfies  the estimate
\[
\|f(t)\|\leq  \|f(t_0)\|+\int_{t_-}^{t_+}\|B(s)\|ds\qquad\forall\,t\in\R
\]
where $t_-:=\min\{t_0,t\}$ and $t_+:=\max\{t_0,t\}$.
\end{Lemma}
\begin{beweis}
(cf. Lemma~2.2  in~\cite{NaSi2})\\
We first prove uniqueness of the solution.
Given two solutions $f_1,\,f_2,$ then for all $\psi\in\cl H.$ we have
\[
(f_1(t)-f_2(t))\psi=\int_{t_0}^t\frac{d}{ds}\big(f_1(s)-f_2(s)\big)\psi \,ds
=\int_{t_0}^t i\big[f_1(s)-f_2(s),A(s)\big]\psi \,ds\,.
\]
By Lemma~\ref{Lemma2} we thus obtain
\[
\|(f_1(t)-f_2(t))\|\leq \int_{t_-}^{t_+} \left\|\big[f_1(s)-f_2(s),A(s)\big]\right\| \,ds
\leq 2M \int_{t_-}^{t_+} \left\|f_1(s)-f_2(s)\right\| \,ds.
\]
Gronwall's Lemma then proves that $f_1(t)-f_2(t)=0$ and hence we have uniqueness.

Next, we prove existence of the solution $f$, and for this we recall the Dyson series for
propagators, cf. Theorem~X.69 in \cite{ReSi2}, where the properties below are proven.
The norm convergent Dyson series is
\[
U(t,t_0):=\un + \sum_{n=0}^\infty i^n\int_{t_0}^t\int_{t_0}^{t_1}\cdots\int_{t_0}^{t_{n-1}}A(t_1)\cdots A(t_n)\,dt_n\cdots dt_1
\]
which is a unitary, and the integrals are defined w.r.t. the strong operator topology. Its basic properties are
$U(t,t)=\un$, $U(t,s)^*=U(s,t)$, $U(r,s)U(s,t)=U(r,t)$ and $U(s,t)$ is strong operator differentiable in both entries.
As $U(t,s)$ satisfies the equations
\[
i\frac{d}{dt}U(t,t_0)\psi=-A(t)U(t,t_0)\psi \qquad\hbox{and}\qquad i\frac{d}{dt}U(t,t_0)^*\psi = U(t,t_0)^*A(t)\psi
\]
for all $\psi\in\cl H.$, it is easy to verify that
\[
f(t):=U(t,t_0)\Big(f_0+\int_{t_0}^tU(s,t_0)^*  B(s)U(s,t_0)\,ds
\Big)U(t,t_0)^*
\]
satisfies
\[
\frac{d}{dt}f(t)\psi = i[f(t),A(t)]\psi+B(t)\psi\quad\forall\,\psi\in\cl H.,\quad
\hbox{and}\quad f(t_0)=f_0\in\al B.(\cl H.).
\]
(We used the fact that the product of strong operator differentiable maps is again strong operator differentiable).
Then
\[
\|f(t)\|\leq\|f_0\|+\Big\| \int_{t_0}^tU(s,t_0)^*  B(s)U(s,t_0)\,ds   \Big\|
\]
and application of Lemma~\ref{Lemma2} to the last integral produces the claimed estimate.
\end{beweis}

\section*{Acknowledgements.}

We wish to thank Professors B. Nachtergaele and R. Sims for clarifying their
work in \cite{NaSi} to us in a number of emails and sending us their manuscript of
 \cite{NaSi2}.


\end{document}